\documentclass[reprint,aps,prx,superscriptaddress,longbibliography,nofootinbib,floatfix]{revtex4-2}
\usepackage{amsmath,amssymb}
\usepackage{graphicx}
\usepackage{array}
\usepackage[hidelinks]{hyperref}
\usepackage{microtype}

\newcommand{\kB}{k_{\mathrm{B}}}
\newcommand{\eps}{\varepsilon}

\makeatletter
\def\active@comma{,}
\newcommand\suppref[1]{\@ifundefined{S@#1}{\textbf{??}\@latex@warning{Label `#1' of the supplement undefined}}{\@nameuse{S@#1}}}
\makeatother
\expandafter\def\csname S@FirstPage\endcsname{}
\expandafter\def\csname S@LastBibItem\endcsname{640}
\expandafter\def\csname S@LastPage\endcsname{}
\expandafter\def\csname S@fig:systems2\endcsname{S2}
\expandafter\def\csname S@fig:systems3\endcsname{S3}
\expandafter\def\csname S@fig:systemsb\endcsname{S1}
\expandafter\def\csname S@sec:inputs\endcsname{S1}
\expandafter\def\csname S@tab:energy\endcsname{S3}
\expandafter\def\csname S@tab:history\endcsname{S2}
\expandafter\def\csname S@tab:systems\endcsname{S1}
}

\begin{document}

\title{A technological capability scale for civilizations from the Kardashev and Barrow scales}

\author{Volkan Gurses}
\affiliation{Department of Electrical Engineering and Computer Science, Massachusetts Institute of Technology, Cambridge, MA, USA}
\affiliation{Research Laboratory of Electronics, Massachusetts Institute of Technology, Cambridge, MA, USA}
\affiliation{Department of Electrical Engineering, California Institute of Technology, Pasadena, CA, USA}
\affiliation{Institute for Quantum Information and Matter, California Institute of Technology, Pasadena, CA, USA}

\begin{abstract}
We introduce a technological capability scale that ranks civilizations by the control operations they can perform per second. A control operation is defined as setting one binary degree of freedom of matter to a specified value, whatever its previous state. Building on constructor theory, we sort them into construction and its special cases on information media, namely processing, storage and transfer. The index $\Xi = K + B$ adds to the Kardashev index $K$ an efficiency index $B$ of energy per operation, measured from the thermal floor at the microwave background temperature. The Kardashev scale gives the ceiling for resets of thermally random bits with heat radiated into the present sky, and the Barrow scale gives the floor, which for such a reset is the thermal floor at the heat-rejection temperature at every level. Since 1950 the energy per logic operation has fallen 11.7 orders of magnitude, around 5 through miniaturization. Humanity, at $K = 0.73$, computes with CMOS logic gates at $B = -0.78$ in primary energy. We describe the ultimate technology, at $\Xi = 7.58$, which would dissipate the conjectured largest power $c^5/4G$ into an uncharged, non-rotating black hole near its largest mass in de Sitter space.
\end{abstract}


\maketitle

\section{Introduction}

Kardashev classified technologically developed civilizations by their energy consumption while estimating how much power an extraterrestrial civilization could put into a radio transmitter \citep{kardashev1964transmission}. Type I has a ``technological level close to the level presently attained on the earth'', Type II commands ``the energy radiated by its own star'' and Type III is ``in possession of energy on the scale of its own galaxy''. He put their consumption at $4\times10^{12}$\,W (his estimate for humanity in 1964), $4\times10^{26}$\,W (the output of the Sun) and $4\times10^{37}$\,W (the output of $10^{11}$ Suns). The reading of Type I as the power a planet receives from its star came later, and the earliest instance that Gray~\citep{gray2020extended} found dates from 1994. Sagan placed the types ten orders of magnitude apart, proposing ``Type 1.0 as a civilization using $10^{16}$ watts for interstellar communication; Type 1.1, $10^{17}$ watts'' and so on \citep{sagan1973cosmic}. We use the form of Gray~\citep{gray2020extended},
\begin{equation}
K = \frac{\log_{10}P - 6}{10},
\label{eq:K}
\end{equation}
with $P$ in watts, so that Types I, II and III correspond to $10^{16}$, $10^{26}$ and $10^{36}$\,W and Kardashev's own values give $K = 0.66$, 2.06 and 3.16. Each type is then a fixed power, independent of a civilization's planet, star and galaxy. World total energy supply was 600.3\,EJ in 2025 \citep{ei2026statistical}, an average of $1.9\times10^{13}$\,W, which gives $K = 0.73$. Sagan classed the civilization of his own day as ``something like Type 0.7''. Gray~\citep{gray2020extended} extended the scale and added companion scales for information, population and mass of constructions, and {\'C}irkovi{\'c}~\citep{cirkovic2015kardashev} reviewed its assumptions.

Barrow proposed an inward counterpart that ranks civilizations by the smallest scale of matter they manipulate \citep{barrow1998impossibility}. A Type I-minus civilization ``is capable of manipulating objects over the scale of themselves'' by ``building structures, mining, joining and breaking solids''. Type II-minus manipulates genes, III-minus molecules and bonds, IV-minus atoms, V-minus nuclei, VI-minus elementary particles and $\Omega$-minus ``the basic structure of space and time''. We call Barrow's types levels, to distinguish them from Kardashev's types. In Barrow's view humanity has ``long been a Type I-minus civilization'', is Type II-minus through genetics ``in several respects'', has ``some Type III-minus abilities'' and has ``only just entered the Type IV-minus domain''. He found it ``struggling to maintain'' V-minus, a status he rested on fission power and nuclear explosions, and ``not yet a Type VI-minus civilization'' \citep[pp.~133--135]{barrow1998impossibility}. {\'C}irkovi{\'c}~\citep{cirkovic2015kardashev} placed humanity at ``about Type II-minus, with some aspirations toward Type III-minus''. We say that a civilization holds a level when it sets a degree of freedom of that level's object to values it has specified. On this criterion humanity has held III-minus since chemists first assembled molecules step by step to a designed structure \citep{fischer1907polypeptiden}, holds II-minus through recombinant deoxyribonucleic acid (DNA) and holds IV-minus in the laboratory, where Eigler \& Schweizer~\citep{eigler1990positioning} positioned single atoms in 1990. Through laboratory control of single nuclear and electron spins \citep{pla2013nuclear,koppens2006driven} humanity holds V-minus and, on our criterion and contrary to Barrow's assessment, VI-minus. Reactors transmute nuclei in bulk and specify no state for each nucleus, so they do not count towards V-minus. We follow Barrow in treating the levels as cumulative.

Dyson paired a power budget with a cost per operation, characterizing life by its rate of entropy production in bits, its dissipated power divided by $\kB T\ln 2$ \citep{dyson1979time}. Bradbury sized a Matrioshka brain at $10^{42}$ operations per second from $10^{26}$\,W at $10^{-16}$\,J per operation \citep{bradbury1999matrioshka,bostrom2003astronomical}, and Sandberg computed Landauer-limited erasure rates for computers at their radiator temperatures \citep{sandberg1999physics}. {\'C}irkovi{\'c} \& Bradbury~\citep{cirkovic2006galactic} applied the Landauer bound $\Delta E/\kB T\ln 2$ to the bits a civilization can process with an energy $\Delta E$ at temperature $T$ and argued that ``we should either entirely abandon or significantly modify'' Kardashev's classification. Wright~\citep{wright2020dyson} computed about $10^{47}$ operations per second for a Dyson sphere at the Landauer limit, and Wright~\citep{wright2023thermodynamics} extended the analysis under a budget of mass. Sharma~\citep{sharma2026cognitive} defined a cognitive Kardashev scale as the product of power, the fraction devoted to computation and the operations per joule of current hardware, and gave humanity's distance from Type I beside that of its hardware from the Landauer bound. Gurovich divided world energy production by the global proof-of-work hashrate to define a state variable $P/H$, in joules per hash, whose floor is a multiple of the Landauer bound at 300\,K \citep{gurovich2026kardashev}. Kempes \textit{et al.}~\citep{kempes2017thermodynamic} found cellular translation to lie about an order of magnitude above the Landauer bound. Sagan and Gray~\citep{gray2020extended} gave stored bits a separate scale beside power, and Hilbert \& L{\'o}pez~\citep{hilbert2011world} measured the world's stored bits and computations. Jiang \& Das~\citep{jiang2025earthbound} combined energy, information, construction and population into one index, and Freitas and Chaisson ranked systems by information rate and energy rate per unit mass \citep{freitas1984xenopsychology,chaisson2011energy}. Vidal combined the two scales into a two-dimensional metric that labels a civilization by a pair such as $\mathrm{K_{II}}$-$\mathrm{B_{\Omega}}$. He attached a length to each Barrow level and read the Barrow scale as a trend towards efficiency \citep{vidal2011black,vidal2014beginning}. {\'C}irkovi{\'c}~\citep{cirkovic2015kardashev} noted that the two scales are ``not entirely decoupled'' and left their interrelation to exploratory engineering, beyond the scope of his review. Vidal \textit{et al.}~\citep{vidal2026technosignatures}, citing Vidal~\citep{vidal2016stellivore}, noted that a civilization may co-develop along both scales. Haqq-Misra \textit{et al.}~\citep{haqqmisra2025projections} read the Kardashev scale as a limit on luminosity and considered technospheres that harvest stellar mass, and Smart~\citep{smart2012transcension} proposed measures of efficiency and density as more appropriate. To our knowledge, no prior work counts operations of every class on matter under one budget, and none assigns a physical least energy per operation to each Barrow level.

Neither scale counts the operations a civilization performs with its power or its devices. Kardashev converted $P$ into bits per second only for radio communications, by dividing the power received from an isotropic transmitter by a noise energy of $100\,\kB T_N$ per bit \citep{kardashev1964transmission}. Barrow gave cost as the reason for the inward direction. Manipulating the large-scale world ``requires huge energy resources'', and manipulation at smaller scales had proved cheaper \citep[p.~133]{barrow1998impossibility}. This paper introduces a scale for technological civilizations built from the Kardashev and Barrow scales. Its quantity is the number of control operations a civilization can perform per second, $\Omega = P/\eps$, and its capability index is $\Xi = K + B$, where $K$ is the Kardashev index and $B$ is an efficiency index. $\Omega$ counts every operation by which a civilization sets matter to a specified state. Kardashev's bit rate and the operation rates cited above are its special cases for information transfer and for processing. In the sense of constructor theory the control operation is the elementary task of construction, and processing, storage and transfer are construction performed on information media. The Kardashev scale gives the ceiling, the power $P$ a civilization commands, and the Barrow scale the floor, the least value of the energy per operation $\eps$ at each level of matter. Barrow and Vidal describe the cost that real devices have reached, which differs from the least cost possible at each level, and reading the Barrow scale as that least cost is our convention. We measure the efficiency index $B$ from the thermal floor at the temperature of the cosmic microwave background, so that $\Xi$ cannot exceed $K$ for operations that erase and radiate their heat into the present sky. We show that descending the levels cannot lower the floor, follow humanity's capability in each class of operation back to the first writing, estimate how much of its gain in processing came from miniaturization, and place the ultimate technology at the corner of the $(K, B)$ plane. Section~\ref{sec:budget} defines the operation and the indices, Section~\ref{sec:levels} gives each level an energy, Section~\ref{sec:results} places humanity, its machines and hypothetical civilizations on the scale, Section~\ref{sec:companions} treats storage and transfer, and Section~\ref{sec:known} states the assumptions and scope.

\section{The operation budget}
\label{sec:budget}

We call a control operation an event that sets one binary degree of freedom of matter, or of any other physical system a civilization can act on, to a value specified in advance, whatever the previous state of that degree of freedom. In the constructor theory of Deutsch~\citep{deutsch2013constructor}, who expressed the laws of physics through the transformations that can and cannot be caused, a control operation is a task, and a device that performs it and remains able to perform it again is a constructor. The specification may be a program, a mask, a template or a design. The degree of freedom belongs to the object of a Barrow level. A positional degree of freedom has the length of the level, while setting an internal one, such as a spin, confines no carrier to that length. After the operation the matter is correlated with the specification, and this correlation is a thermodynamic resource \citep{parrondo2015thermodynamics}. We count operations by the entropy removed, in bits, so a choice among $W$ equally likely alternatives counts as $\log_2 W$. A logic gate that overwrites its output performs one operation, a device that draws an atom from a vapour and places it at a specified site performs as many operations as the placement removes bits of entropy, and a machine that adds one of $a$ kinds of monomer against a template performs $\log_2 a$, two for a polymerase and $\log_2 20 = 4.32$ for a ribosome. In the tables an atom placed counts one operation, a lower bound. Copying a template, like exposing a resist through a mask, is logically reversible \citep{bennett1982thermodynamics}, and we count its bits as operations by convention. A monomer drawn from a mixture in which its kind has mole fraction $x$ needs $\kB T\ln(1/x)$ more work than one drawn from a pure supply at the same concentration, $\kB T\ln a$ for an equimolar pool of $a$ kinds, and this work, which matches the count $\log_2 a$, is stored in the product relative to the pool and is not heat. Heating, mixing, combustion and propulsion change matter without setting it to specified values, and we do not count them. Nor do we count moving an object between two known places, which can be undone, or ammonia synthesis, which produces an order fixed by the reaction without a specification.

A civilization that runs $N$ devices, each performing $f$ operations per second at an energy $\eps$ per operation, spends $Nf\eps$ watts in steady state, as heat or as energy stored in its products. This cannot exceed the power it commands, so the number of operations it can perform per second is at most
\begin{equation}
\Omega = \frac{P}{\eps},
\label{eq:Omega}
\end{equation}
with $P$ in watts and $\eps$ in joules per operation. The number and speed of the devices do not appear in $\Omega$, which is therefore independent of the architecture for a device-level unit of counting, as for Lloyd's bound \citep{lloyd2000ultimate}. That bound counts the reversible steps of a given mass, whereas $\Omega$ counts the irreversible operations a given power sustains.

Each node that a logic gate overwrites is one operation, so the unit of counting is the switching event, and an addition or a floating-point operation consists of many. Where the switching events of a technique are not documented, we count each bit of its result as one operation, a lower bound on the operations it performs. An addition of $n$ decimal digits is then $n\log_2 10$ operations (33 for ten digits), a $w$-bit addition or instruction $w$, a 64-bit floating-point operation 64, a 16-bit one 16 and a SHA-256d hash 256. Switching events, qubit resets, synaptic events and bits written count one each. This count is an upper bound on the entropy removed, and a counted operation reaches the floor of Section~\ref{sec:levels} only when the prior state of its degree of freedom is equiprobable and uncorrelated with the controller. We take the switching event of the International Roadmap for Devices and Systems (IRDS) of the Institute of Electrical and Electronics Engineers, with an energy $CV^2$ of 0.65\,fJ in 2023 for a high-performance gate driving three others \citep{irds2023moremoore}. Published energies per switching event span two orders of magnitude, from 32\,aJ for a minimum-width transistor at the 14\,nm node \citep{lee2017cmosdata} to about 3\,fJ for a practical circuit with its wiring \citep{theis2017end}, and we give results that depend on the choice among them as ranges. A 32-bit integer addition costs about 3\,fJ per operation at 45\,nm \citep{horowitz2014computing} and 0.9\,fJ at 7\,nm \citep{jouppi2021ten}. El~Capitan, which led the TOP500 list through 2025, drew 29.7\,MW while performing $1.81\times10^{18}$ 64-bit floating-point operations per second in the Linpack benchmark \citep{top500nov2025}, or $2.6\times10^{-13}$\,J per operation for the whole machine, the energy of about 400 switching events.

The budget $\Omega$ is a capacity, the rate reached if all of $P$ went to control operations at the stated cost, in the way the Kardashev scale counts every watt regardless of its use. Power delivered as work or heat to move, heat and transform matter counts in the numerator without adding operations. We sort control operations into four classes by what becomes of the specified bit. Deutsch~\citep{deutsch2013constructor} generalized computation to every physical transformation, and we follow him in taking construction as the general class. The operations of construction on information media \citep{deutsch2015constructor} form three classes, in which processing overwrites the bit after use, storage holds it in place and transfer recreates it at another place. These are the computation, storage and communications whose worldwide capacities Hilbert \& L{\'o}pez~\citep{hilbert2011world} measured. The other operations, which build the bit into the matter of a product as fabrication and synthesis do and as von Neumann's universal constructor would \citep{vonneumann1966theory}, remain in construction. We place each system in the class of the operation by which we count it, so a system counted by operations of every class, such as the programmable quantum field of Section~\ref{sec:eta}, belongs to construction. Since operations of different classes are not interchangeable, a capacity refers to one class, and the largest over classes is the civilization's overall capacity. The realized rate is lower. Data centres used 1.75\,EJ of electricity in 2025 \citep{iea2026keyquestions}, an average of 55\,GW, or about 130\,GW of primary energy at the factor given below. Their primary energy, part of which goes to memory, networking and cooling, is 0.7\% of humanity's total energy supply and gives $K = 0.51$.

We take $P$ to be the power a civilization's technology draws from its environment and count $\eps$ at the same boundary, as the energy drawn per operation, including the conversion of fuel into electricity and its delivery. For humanity this boundary is that of primary energy, and for electricity in 2025 we take 2.3 units of primary energy per unit delivered. The factor follows from the world generation mix \citep{ember2026global}, efficiencies of 38\% and 33\% for fossil and nuclear plants and an assumed 30\% for bioenergy plants (electronic supplementary material, section~\suppref{sec:inputs}). Other renewables are counted at their output as in the physical-energy-content accounting of the total energy supply \citep{ei2026statistical}. Transmission and distribution losses are 6.5\% \citep{worldbank2026losses}. Over plausible efficiencies the factor is 2.2 to 2.5. For 1950 we take 3.5, with a range of 3 to 5, from an efficiency of 24\% for United States fossil steam plants, a hydropower share near 30\% and losses near 13\% \citep{eia2012aer}. We apply the same factor to earlier electricity, for which it is a lower bound. For enzymes we charge the heating value of the glucose from which a cell makes its adenosine triphosphate (ATP), and the energy of synthesizing the monomers. The heating value is 2.8\,MJ/mol \citep{nist2026webbookglucose} for about 30 ATP \citep{milo2015cell}, or 1.55 times the free energy of ATP hydrolysis in the cell. We count the glucose as biomass is counted in the total energy supply, leaving out the fossil energy spent to grow and refine it. The floors of Section~\ref{sec:levels} bound the heat released at the device and hence also the energy drawn per operation, so the maturity gap defined below includes the losses of conversion, and we give values at the device where they matter.

Because $K$ is a logarithm of power in steps of ten orders of magnitude, we index $\Omega$ with the same logarithm, normalized to the thermal floor of Section~\ref{sec:levels} at a reference temperature $T_0 = 2.725$\,K, the present temperature of the cosmic microwave background \citep{fixsen2009temperature}. At the present epoch the background sets the temperature of the sky, the coldest sink that surrounds every civilization, and every civilization at rest with respect to the background measures the same temperature. We fix $T_0$ at this value, so that it is a unit, as $10^{6}$\,W is in Eq.~\eqref{eq:K}. The efficiency index and the capability index are
\begin{equation}
B = \frac{1}{10}\log_{10}\frac{\eps_{T_0}}{\eps}, \qquad \eps_{T_0} = \kB T_0\ln 2 = 2.61\times10^{-23}\,\mathrm{J},
\label{eq:B}
\end{equation}
\begin{equation}
\Xi = K + B = \frac{1}{10}\log_{10}\frac{\Omega}{\Omega_0},
\label{eq:Xi}
\end{equation}
where $\Omega_0 = 10^{6}\,\mathrm{W}/\eps_{T_0} = 3.8\times10^{28}$\,s$^{-1}$ is the budget of a civilization of $10^{6}$\,W on the thermal floor at $T_0$. The unit of time cancels in $\Xi$, and the scale has two references, namely the power $10^{6}$\,W and the temperature $T_0$. A civilization whose operations cost the thermal floor at $T_0$ has $\Xi = K$, and a negative $B$ measures in Kardashev units how far its operations lie above that floor. Because $K$ and $B$ have equal weight, $\Xi$ is linear in $\log_{10}\Omega$, ten orders of magnitude in efficiency or in power raise $\Xi$ by one, and lines of constant $\Xi$ on the $(K, B)$ plane have slope $-1$. Another reference temperature $T_0'$ shifts every $B$ by $\tfrac{1}{10}\log_{10}(T_0'/T_0)$. The primary-energy factor for electricity in 2025 lowers $B$ by 0.034 to 0.040 below its value at the device.

The floor $\eps_{\mathrm{floor}}$ of an operation is its least energy at $T_0$, derived in Section~\ref{sec:levels} for the operation and carrier assigned to its level. For heat radiated into the present sky the largest $B$ possible for the operation is $B_{\mathrm{floor}} = \tfrac{1}{10}\log_{10}(\eps_{T_0}/\eps_{\mathrm{floor}}) \le 0$, which depends only on the operation and on physics and is zero wherever the floor is the thermal floor. An operation on the thermal floor at a rejection temperature $T$ has $B = \tfrac{1}{10}\log_{10}(T_0/T)$, zero at $T_0$ and $-0.20$ at 300\,K. A radiator rejects net heat into the present sky only while it is warmer than $T_0$, so when entropy leaves as heat radiated into the sky, $B \le B_{\mathrm{floor}} \le 0$ and $\Xi \le K$ for operations that remove $\kB\ln 2$ of entropy per counted bit. Only a new sink for entropy (Eq.~\eqref{eq:OmegaS}) or a sink colder than the microwave background, such as a black hole or, in the far future, the de Sitter horizon of Section~\ref{sec:known}, could raise $\Xi$ above $K$. The floor of a level is the least over the operations of that level, and since the levels are cumulative, the floor of a civilization is the least over the levels it holds. The maturity gap $\mu = B - B_{\mathrm{floor}}$, at most zero for heat radiated into the present sky, is the distance of an operation the civilization performs from the floor at $T_0$. It includes a term $\tfrac{1}{10}\log_{10}(T_0/T)$ from the temperature $T$ at which the civilization rejects its heat, and the distance of its devices from the thermal floor at that temperature. We describe a civilization, for each class of operation, by $(K, B_{\mathrm{floor}}, \mu)$. The quantities $P$, $\eps$ and hence $K$, $B$ and $\Xi$ are measured for a real civilization, the primary-energy factors are estimates, and the reference temperature $T_0$, the reliability reference of Section~\ref{sec:levels} and the length, operation and carrier assigned to each level are conventions. We derive the floors and the limits of Section~\ref{sec:known} from physics as known in 2026, which we denote $\mathcal{L}$. It consists of the second law of thermodynamics and its extension to horizons, linear and unitary quantum mechanics, general relativity with a positive cosmological constant, and the Standard Model with massive neutrinos, together with dark matter of unknown nature. A discovery that changes $\mathcal{L}$ moves $B_{\mathrm{floor}}$ and $\mu$ by equal and opposite amounts and leaves $B$ unchanged.

\section{Energy per operation at each Barrow level}
\label{sec:levels}

We take the least energy of an operation to be the larger of two terms. The quantum term is the energy an operation must concentrate on a carrier to confine it to the length of the level. It depends on that length and on the mass of the carrier and contains no temperature. The thermal term arises because an operation lowers the entropy of the matter it sets, and in steady state that entropy leaves as heat to a bath at temperature $T$. Reversible logic costs $\kB T\ln 2$ for each bit erased, as errors or outputs, plus an adiabatic loss that falls as the logic slows, and has no minimum dissipation per logical operation \citep{bennett1982thermodynamics,athas1994low}. A switch that discharges its barrier at every operation turns its confinement energy into heat \citep{zhirnov2003limits}, whereas a placed atom or nucleon keeps its confinement energy in the depth of its site. We therefore count the quantum term as dissipated only for a switch reset at every operation, and we call a confinement energy entered as the cost of an operation a model value. A model value bounds the heat of an operation only if the operation does not recover that energy.

\subsection{The quantum term}

We take the energy of confining a carrier of mass $m$ to a length $l$ to be
\begin{equation}
E_{\mathrm{loc}}(l, m) = \sqrt{\left(\frac{\hbar c}{l}\right)^2 + \left(mc^2\right)^2} - mc^2,
\label{eq:Eloc}
\end{equation}
which is $\hbar^2/2ml^2$ above the reduced Compton wavelength of the carrier and $\hbar c/l$ below it. It is the kinetic energy of a momentum $\hbar/l$, and we adopt Eq.~\eqref{eq:Eloc} as a convention. A state confined by hard walls to a region of length $l$ has at least $\pi^2$ times the nonrelativistic value, and a state whose standard deviation of position is $l$ at least a quarter of it. For an electron switch Eq.~\eqref{eq:Eloc} is the barrier that Zhirnov \textit{et al.}~\citep{zhirnov2003limits} estimated from the Heisenberg relation, at which a switch of width $l$ loses distinguishability to tunnelling. It is 0.61\,aJ at 0.1\,nm and 6.1\,zJ at 1\,nm, and for a hydrogen atom at 0.1\,nm it is $3.3\times10^{-22}$\,J, 13 times $\kB T_0\ln 2$. The Margolus-Levitin theorem bounds the rate of operations of a device holding an energy $E$ by $2E/\pi\hbar$ \citep{margolus1998maximum}, but a reversible device can recover that energy, so the bound sets no floor on cost \citep{lloyd2000ultimate}.

The thermal term is the long-length limit of the isothermal work to confine a particle to one of two regions of length $l$. This work is $\kB T\ln 2$ when $l$ is much longer than the thermal de Broglie wavelength of the particle and approaches $3\pi^2/4 \approx 7.4$ times the quantum term for hard walls when $l$ is much shorter. At $T_0$ the wavelength is 45\,nm for an electron and 1.1\,nm for a hydrogen atom. At any length the least heat released is at most $1.12\,\kB T\ln 2$ for hard walls, and the rest of the work is stored in the confined state, so that in the short limit almost none reaches the bath. Numerical values are given in the electronic supplementary material, section~\suppref{sec:inputs}.

\subsection{The thermal term}
\label{sec:thermal}

Erasing a bit releases on average at least $\kB T\ln 2$ of heat, 2.9\,zJ at 300\,K and 26\,yJ at $T_0$ \citep{landauer1961irreversibility}. Any logically irreversible manipulation of information, whatever its carrier, must increase the entropy of degrees of freedom that bear no information \citep{bennett2003notes}. The least work to reset a system is its change in non-equilibrium free energy, $\kB T\ln 2$ for a random bit \citep{parrondo2015thermodynamics}. The simplest operation on matter, confining one molecule to one half of a box, runs Szilard's engine in reverse and costs this amount \citep{szilard1929entropieverminderung,bennett1982thermodynamics}. B{\'e}rut \textit{et al.}~\citep{berut2012experimental} placed a colloidal bead in one of two wells, itself a control operation on matter, and approached the bound as they slowed the operation. We call the bound the thermal floor. With unequal prior probabilities the least heat is $T$ times the entropy removed. A controller that first measures the prior state can set it without dissipation, and the cost moves to erasing its memory of the measurement \citep{bennett1982thermodynamics}. Vaccaro \& Barnett~\citep{vaccaro2011information} showed that erasure into a reservoir of spins can cost angular momentum in place of energy.

A polymerase copying a template can, like a reversible computation \citep{bennett1973logical}, run backward step by step near equilibrium and return each monomer to its pool, so copying has no minimum dissipation per monomer \citep{bennett1982thermodynamics}. The least heat that construction releases in steady state, $T$ times the entropy it removes from the matter, is set by the chemistry and need not reach $\kB T\ln 2$ per specified bit, and writing a pattern into a medium prepared in a known state removes no entropy from the medium. The thermal floor therefore bounds every reset in processing, setting a spin, placing an atom from a vapour and any other operation that sets a degree of freedom whose prior state was thermally random, and it does not bound copying a design into prepared matter. When the states of the matter differ in energy, part of the work may remain in the product as the free energy of a reaction or of its bonds. That stored work can in principle be recovered and is not part of the floor. Where the feedstock supplies the free energy, as in crystallization, we count that free energy in $P$.

For a degree of freedom that is thermally random and uncorrelated with the controller, an operation lowers its entropy by $\kB\ln 2$, and the second law requires at least this entropy to be carried away, so
\begin{equation}
\Omega \le \frac{\dot S}{\kB\ln 2},
\label{eq:OmegaS}
\end{equation}
where $\dot S$ is the rate at which the civilization can dispose of entropy into a heat sink, into a reservoir of another conserved quantity \citep{vaccaro2011information,guryanova2016thermodynamics} or into a stock of low-entropy matter \citep{bennett2019comment}. Degrading work at a rate $P$ into heat rejected at $T$ disposes of $\dot S = P/T$, and Eq.~\eqref{eq:OmegaS} reduces to Eq.~\eqref{eq:Omega} on the thermal floor. Wall~\citep{wall2012proof} proved the generalized second law, which extends Eq.~\eqref{eq:OmegaS} to horizons, for free semiclassical quantum fields crossing a causal horizon. Equation~\eqref{eq:OmegaS} depends on physics only through the sources and sinks that set $\dot S$. A discovery of new sources of work or new sinks for entropy can raise the budget, and removing the entropy cost of an operation would violate the second law.

The temperature $T$ is that at which the civilization finally rejects its heat, since a bit erased at $T_c$ whose heat is pumped to $T_h$ costs at least $\kB T_h\ln 2$ from the supply, independent of $T_c$, and a confinement energy dissipated at $T_c$ costs $(T_h/T_c)E_{\mathrm{loc}}$. A radiator at rest with respect to the microwave background rejects net heat into the present sky only while it is warmer than the background (Section~\ref{sec:budget}), so for that heat the thermal floor at $T_0$ is an infimum.

A structure that must persist for a time $t$ against thermal hopping needs a barrier of about $\kB T\ln(\nu_0 t)$, where the attempt frequency $\nu_0 \approx 10^{13}$\,s$^{-1}$ is set by atomic vibrations. The barrier is $30\,\kB T$ for one second and $71\,\kB T$ for the Hubble time $1/H_0 = 4.6\times10^{17}$\,s \citep{planck2020cosmological}. It is held and need not be dissipated, while a step committed passively dissipates the free energy that drives it. Since a step that dissipates $\eps$ runs forward $e^{\eps/\kB T}$ times as often as backward \citep{bennett1982thermodynamics}, leaving a fraction $p$ undone costs about $\kB T\ln(1/p)$. Salehi Fashami \textit{et al.}~\citep{salehifashami2013energy} obtain about $2\kB T\ln(1/p)$ for fault-tolerant switching. A switch that must complete $N$ operations without a thermal error needs a barrier of about $\kB T\ln N$, and a switch that runs at $\nu_0$ for a Hubble time completes $\nu_0/H_0 = 4.6\times10^{30}$ operations, so the two criteria coincide for such a switch. As a reliability reference we use $\kB T\ln(\nu_0/H_0) = 71\,\kB T$, which lies 2.0 orders of magnitude above the thermal floor at any temperature and is 0.29\,aJ at 300\,K. At $T_0$ thermal hopping governs only barriers of curvature frequency below ${3.6\times10^{11}}$\,s$^{-1}$, and one of curvature frequency $\nu_0$ is crossed by tunnelling \citep{hanggi1990reaction}, so there we keep the reference as a convention (electronic supplementary material, section~\suppref{sec:inputs}). The Hubble time, like $T_0$, is the same for every civilization at the present epoch. A switch driven slowly enough reaches any reliability at a dissipation of order $\kB T\ln 2$ \citep{keyes1970minimal}, so the reference is not a bound. A wrong monomer incorporated, unlike a step left undone, is an error set by the difference in binding free energy between right and wrong monomers, which is held and not dissipated, and by proofreading, which spends more free energy per correct step \citep{hopfield1974kinetic}.

For an electron the quantum term equals the thermal floor at $l = \hbar/\sqrt{2m_e\kB T\ln 2}$, which is 15\,nm at $T_0$ and 1.5\,nm at 300\,K, the minimum switch size of Zhirnov \textit{et al.}~\citep{zhirnov2003limits}. It equals the reliability reference at 1.5\,nm at $T_0$ and 0.14\,nm at 300\,K. Above these lengths the thermal term dominates, and for atoms heavier than 12.8\,u it does so down to the atom. A massless carrier costs $\hbar c/l$, which crosses the thermal floor at 1.2\,mm at $T_0$ and 11\,$\mu$m at 300\,K, so a photonic device that absorbs its photons lies above the thermal floor at $T_0$ at every level below I-minus. The energy per operation is
\begin{equation}
\eps(l, m, T) = \max\left\{E_{\mathrm{loc}}(l, m),\; \kB T \ln N_{\mathrm{ops}}\right\},
\label{eq:eps}
\end{equation}
with $N_{\mathrm{ops}} = 2$ for the thermal floor and $N_{\mathrm{ops}} = \nu_0/H_0$, giving $71\,\kB T$, for the reliability reference.

\begin{table*}[t]
\caption{Least energy per bit of a control operation at each of Barrow's levels, from Eq.~\eqref{eq:eps} on the thermal floor with device and sink at $T_0$. The object is Barrow's, and the example operation, length $l$ and carrier are our convention. Each entry that is not a model value is the least energy of an operation that sets a degree of freedom whose prior state was thermally random (Section~\ref{sec:thermal}). For I-minus $l$ is the least length at which every carrier, massless included, has only the thermal term. A placed atom keeps its confinement energy in its site, so the IV-minus entry is the thermal floor for every atom. The last column gives orders of magnitude above $\kB T_0\ln 2 = 2.6\times10^{-23}$\,J, or $-10B_{\mathrm{floor}}$. The lower row at IV-minus, V-minus and VI-minus is a model value. At V-minus and VI-minus the upper row is the cheapest operation, setting a spin, and the model value bounds the lower-row operation only if its stored energy is not recovered. For the electron at IV-minus the value in parentheses is the exact isothermal work of confinement with hard walls, almost all of which is stored in the confined state. At $\Omega$-minus, where we identify no control operation, the entry is the confinement energy at the Planck length, also a model value.}
\label{tab:levels}
\begin{ruledtabular}
\begin{tabular}{lllllll}
Level & Object & Example operation & $l$ (m) & Carrier & $\eps$ (J) & \begin{tabular}[b]{@{}l@{}}Orders above\\ $\kB T_0\ln 2$\end{tabular} \\
\hline
I-minus & objects of own scale & set a switch or position & $\ge 1.2\times10^{-3}$ & any & $2.6\times10^{-23}$ & $0$ \\
II-minus & gene & add a monomer & $2\times10^{-9}$ & monomer & $2.6\times10^{-23}$ & $0$ \\
III-minus & molecule, bond & make or break a bond & $10^{-9}$ & molecule & $2.6\times10^{-23}$ & $0$ \\
IV-minus & atom & place an atom & $10^{-10}$ & atom & $2.6\times10^{-23}$ & $0$ \\
 & & switch an electron & $10^{-10}$ & electron & $6.1\times10^{-19}$ ($4.5\times10^{-18}$) & $4.4$ ($5.2$) \\
V-minus & nucleus & set a nuclear spin & $10^{-15}$ & nucleus & $2.6\times10^{-23}$ & $0$ \\
 & & add a nucleon & $10^{-15}$ & nucleon & $3.3\times10^{-12}$ & $11.1$ \\
VI-minus & elementary particle & set an electron spin & $10^{-18}$ & electron & $2.6\times10^{-23}$ & $0$ \\
 & & create a particle & $10^{-18}$ & massless & $3.2\times10^{-8}$ & $15.1$ \\
$\Omega$-minus & spacetime & none identified & $1.6\times10^{-35}$ & massless & $2.0\times10^{9}$ & $31.9$ \\
\end{tabular}
\end{ruledtabular}
\end{table*}

\begin{figure*}[t]
\centering
\includegraphics[width=0.76\textwidth]{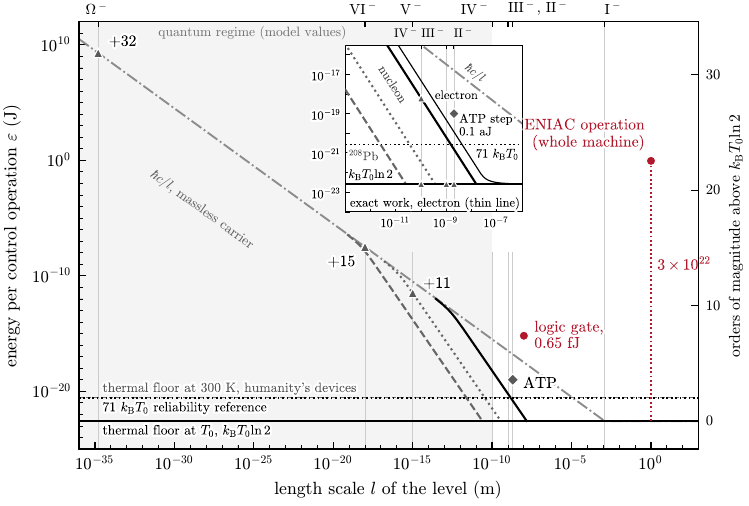}
\caption{Energy per control operation $\eps$, Eq.~\eqref{eq:eps}, against length $l$ at $T_0$ on the thermal floor. Solid, dashed and dotted curves are for an electron, a $^{208}$Pb atom and a nucleon, and every lighter atom lies between the last two. The dash-dotted line is $\hbar c/l$ for a massless carrier. Horizontal lines mark the thermal floor $\kB T_0\ln 2$ (solid), the $71\,\kB T_0$ reliability reference (dotted) and the thermal floor at 300\,K (thin grey dashed), 8\% above the reliability reference. The top axis marks the length of each level (Table~\ref{tab:levels}), placing I-minus at its least length, 1.2\,mm, and the right axis gives orders of magnitude above $\kB T_0\ln 2$. The shaded region below the atom holds model values, and triangles mark those of V-minus, VI-minus and $\Omega$-minus at $+11$, $+15$ and $+32$ orders above the floor. The inset enlarges IV-minus to II-minus with the exact isothermal work of confinement of an electron (thin line), and all its entries but the electron switch lie on the floor. Red circles are humanity's costs of processing at 300\,K, namely an operation of ENIAC for the whole machine and a logic gate at 0.65\,fJ. The dotted red line spans the factor of $3\times10^{22}$ between ENIAC and the floor. The diamond is one ATP hydrolysis step, 0.1\,aJ. Measured points are plotted at nominal lengths of 1\,m, 10\,nm and 2\,nm.}
\label{fig:levels}
\end{figure*}

\subsection{Energies of the Barrow levels}

To apply Eq.~\eqref{eq:eps} to a level we assign it a length and a carrier. Barrow's definitions name objects and give no lengths, and Vidal~\citep{vidal2011black} tabulated lengths without carriers. Table~\ref{tab:levels} lists our assignments, and Fig.~\ref{fig:levels} shows $\eps$ as a function of $l$. Down to the atom the carrier is the matter of the object Barrow names, and the electron at IV-minus stands for a switch of atomic size. Barrow defines I-minus by objects on the scale of a civilization's own members, so we assign it every length above $\hbar c/\kB T_0\ln 2 = 1.2$\,mm, where any carrier, massless included, has only the thermal term. Our lengths depart from Vidal's at I-minus, II-minus and IV-minus, where he gives 1\,m, $10^{-7}$\,m for genes and $10^{-11}$\,m for atoms.

At I-minus, where an operation sets the position or state of a macroscopic object, real devices cost far more than the thermal term. ENIAC, the first electronic reconfigurable general-purpose computer, drew 150\,kW for 5000 ten-digit additions per second \citep{wardept1946eniac}. For the whole machine this is about 0.9\,J per operation, $3\times10^{20}$ times the thermal floor at 300\,K and $3\times10^{22}$ times that at $T_0$. Setting a monomer (II-minus), a bond (III-minus) or the site of an atom 0.1\,nm across (IV-minus) has the thermal floor, 26\,yJ per bit at $T_0$, when the prior state was thermally random. An electron switch of atomic size has the model value 0.61\,aJ, 2.1 times the reliability reference at 300\,K, and tunnels through its barrier with a probability of $e^{-2}$ per attempt, so it is not reliable. The exact work of confinement gives 4.5\,aJ.

At V-minus the cheapest operation, setting a nuclear spin as initializing a nuclear spin qubit does, leaves the nucleons in place, stores a Zeeman energy far below $\kB T$ and has the thermal floor. The costliest example operation adds or removes a nucleon, and we keep its confinement energy at 1\,fm, 21\,MeV or 3.3\,pJ, as a model value. The strong force supplies this energy and the nucleus stores it, so the model value is not a floor on the heat a nuclear operation dissipates. Fusion and fission release energy. In practice freeing a nucleon costs more than the model value, about 30\,MeV of heat per neutron in a spallation source \citep{klein1994spallation}. At VI-minus, setting the spin of an electron or the polarization of a photon has the thermal floor. Creating a particle localized to $10^{-18}$\,m needs a momentum transfer of about 200\,GeV/$c$ and, for a massless quantum, the energy $\hbar c/l$, 32\,nJ, which lies near the electroweak scale set by the masses of the W, Z and Higgs bosons. The energy is a model value, carried by the particle and not dissipated. We know of no binary degree of freedom of spacetime that a civilization could set, and we list the Planck energy, $2\times10^{9}$\,J, at $\Omega$-minus only as the energy needed to localize a quantum at the Planck length.

The floor of every level is therefore the thermal floor for operations that reset a thermally random degree of freedom, and it is independent of length and carrier because the entropy of such an operation, $\kB\ln 2$ per bit, is fixed by counting states. I-minus has the thermal floor, so descending the Barrow levels cannot lower it. Only the quantum term depends on the level, and for some operations it exceeds the thermal floor. The model values for the electron switch of atomic size, the nucleon and the particle have $B_{\mathrm{floor}} = -0.44$, $-1.11$ and $-1.51$, and the Planck energy lies 32 orders of magnitude above the floor at $T_0$.

Biological molecular machines perform the best-measured II-minus and III-minus operations, and humanity has directed them with designed sequences since Itakura \textit{et al.}~\citep{itakura1977expression} and Goeddel \textit{et al.}~\citep{goeddel1979expression} made bacteria produce somatostatin and human insulin from chemically synthesized genes. We take $10^{-19}$\,J, or $24\,\kB T$ at 300\,K, per ATP hydrolysis, the value implied by the $10^{-12}$\,W power of an \textit{E.~coli} cell and its $10^{7}$ ATP molecules hydrolysed per second \citep{milo2015cell}. A ribosome spends four nucleoside triphosphate equivalents per amino acid and a polymerase two per nucleotide \citep{lynch2015bioenergetic}. A ribosome executing a messenger ribonucleic acid (mRNA) thus spends $9.3\times10^{-20}$\,J, or $22\,\kB T$, per specified bit, and a polymerase copying a template $10^{-19}$\,J, or $24\,\kB T$. Bennett~\citep{bennett1982thermodynamics} estimated about $20\,\kB T$ per nucleotide for RNA polymerase, or $10\,\kB T$ per bit. At 300\,K these costs are 32 to 35 times the thermal floor and Bennett's estimate 14 times, all below the reliability reference and of the order of magnitude Kempes \textit{et al.}~\citep{kempes2017thermodynamic} found. Counting the synthesis of the monomers, about 29 phosphate bonds per amino acid and 48 per ribonucleotide in all against the 4 and 2 of chain elongation alone \citep{lynch2015bioenergetic}, raises them by a factor of about 7 for the ribosome and 24 for the RNA polymerase.

\section{Results}
\label{sec:results}

\begin{table*}[t]
\caption{Humanity and hypothetical civilizations on the scale. Each row is a capacity, the stated power charged to one class of operation. Humanity's rows are at its power in the stated year and, before 1950, at the cheapest technique of the class found at or before that year. Machines, technologies and organisms at their own power are in Fig.~\ref{fig:systems} and in the electronic supplementary material, Figs.~\suppref{fig:systemsb} to \suppref{fig:systems3} and Table~\suppref{tab:systems}. $P$ is the power commanded, $\eps$ the energy per control operation of the stated device or floor, $\Omega = P/\eps$, and $K$, $B$ and $\Xi$ follow Eqs.~\eqref{eq:K}, \eqref{eq:B} and \eqref{eq:Xi}, with $\Xi$ computed before rounding. Humanity's power is its total energy supply, 600.3\,EJ in 2025 \citep{ei2026statistical} and 100.7\,EJ in 1950 \citep{smil2017energy,owid2026primaryenergy}, and before 1950 its use of energy including food and feed (Table~\suppref{tab:energy}). Costs for humanity are primary energy per operation, counted per bit of a result for processing (Section~\ref{sec:budget}). We charge electricity at 2.3 units of primary energy per unit delivered in 2025 and 3.5 up to 1950, manual techniques at the metabolic power of their operators, and enzymes for the glucose behind their ATP and monomers. Sources for the costs are in the text and in Tables~\suppref{tab:systems} and \suppref{tab:history}. The rows for 1\,CE, 1455 and 1804 are proxies, at the rate of a 1946 abacus champion, the rate set for the wooden press in 1573 and a loom speed of 1910. The processing row for 1950 is ENIAC, although the Comptometer, EDSAC, SEAC and Whirlwind~I cost less per operation. Humanity's rows since 1950 are at 300\,K, and for the hypothetical rows we charge all of $P$ at the stated device cost or floor, without losses of conversion. The fabrication row is an upper bound on $\eps$ per specified bit, and the quantum computer row is a projection (Section~\ref{sec:gap}). The construction rows are compared with the thermal floor, which does not bound copying a template (Section~\ref{sec:levels}). The Matrioshka brain row is Bradbury's estimate at the device cost of Drexler's rod-logic nanocomputer. The ultimate technology, the conjectured largest power of one system that drops its entropy into an uncharged, non-rotating black hole near the largest mass possible for it in de Sitter space, is the corner of the plane, which no system reaches (Section~\ref{sec:eta}).}
\label{tab:civ}
\begin{ruledtabular}
\begin{tabular}{llllrrr}
 & $P$ (W) & $\eps$ (J) & $\Omega$ (s$^{-1}$) & $K$ & $B$ & $\Xi$ \\
\hline
\multicolumn{7}{l}{\textit{Humanity before 1950, with food and feed}} \\
1\,CE, processing, abacus, per operation (proxy) & $8.8\times10^{10}$ & $12$ & $7.2\times10^{9}$ & $0.49$ & $-2.37$ & $-1.87$ \\
793, storage, scribe, per bit & $1.1\times10^{11}$ & $29$ & $3.9\times10^{9}$ & $0.51$ & $-2.40$ & $-1.90$ \\
1455, storage, Gutenberg press, per bit (proxy) & $2.0\times10^{11}$ & $0.22$ & $8.9\times10^{11}$ & $0.53$ & $-2.19$ & $-1.66$ \\
1795, transfer, Chappe telegraph, per bit & $6.0\times10^{11}$ & $1.7\times10^{3}$ & $3.4\times10^{8}$ & $0.58$ & $-2.58$ & $-2.00$ \\
1804, construction, Jacquard loom, per warp thread (proxy) & $6.3\times10^{11}$ & $0.61$ & $1.0\times10^{12}$ & $0.58$ & $-2.24$ & $-1.66$ \\
1856, processing, arithmometer, per operation & $9.3\times10^{11}$ & $3.8$ & $2.4\times10^{11}$ & $0.60$ & $-2.32$ & $-1.72$ \\
1860, transfer, printing telegraph, per bit & $9.6\times10^{11}$ & $8.8$ & $1.1\times10^{11}$ & $0.60$ & $-2.35$ & $-1.75$ \\
1913, processing, Comptometer, per operation & $2.3\times10^{12}$ & $2.2$ & $1.0\times10^{12}$ & $0.64$ & $-2.29$ & $-1.66$ \\
1929, storage, rotary press, per bit & $2.8\times10^{12}$ & $4.2\times10^{-3}$ & $6.6\times10^{14}$ & $0.64$ & $-2.02$ & $-1.38$ \\
\multicolumn{7}{l}{\textit{Humanity since 1950, total energy supply}} \\
1950, processing, ENIAC, per operation & $3.2\times10^{12}$ & $3.2$ & $1.0\times10^{12}$ & $0.65$ & $-2.31$ & $-1.66$ \\
2025, processing, El~Capitan, per operation & $1.9\times10^{13}$ & $5.9\times10^{-13}$ & $3.2\times10^{25}$ & $0.73$ & $-1.04$ & $-0.31$ \\
2025, processing, 32-bit addition, per operation & $1.9\times10^{13}$ & $2.2\times10^{-15}$ & $8.8\times10^{27}$ & $0.73$ & $-0.79$ & $-0.06$ \\
2025, processing, gate switching & $1.9\times10^{13}$ & $1.5\times10^{-15}$ & $1.3\times10^{28}$ & $0.73$ & $-0.78$ & $-0.05$ \\
2025, processing, projected quantum computer, per reset & $1.9\times10^{13}$ & $2.9\times10^{-5}$ & $6.6\times10^{17}$ & $0.73$ & $-1.80$ & $-1.08$ \\
2025, storage, solid-state drive, per bit & $1.9\times10^{13}$ & $1.0\times10^{-9}$ & $1.9\times10^{22}$ & $0.73$ & $-1.36$ & $-0.63$ \\
2025, transfer, optical module, per bit & $1.9\times10^{13}$ & $4.9\times10^{-11}$ & $3.9\times10^{23}$ & $0.73$ & $-1.23$ & $-0.50$ \\
2025, construction, ribosome on designed mRNA, per bit & $1.9\times10^{13}$ & $1.0\times10^{-18}$ & $1.8\times10^{31}$ & $0.73$ & $-0.46$ & $0.27$ \\
2025, construction, polymerase on designed template, per bit & $1.9\times10^{13}$ & $3.7\times10^{-18}$ & $5.1\times10^{30}$ & $0.73$ & $-0.52$ & $0.21$ \\
2025, construction, chip fabrication, per transistor & $1.9\times10^{13}$ & $1.7\times10^{-3}$ & $1.1\times10^{16}$ & $0.73$ & $-1.98$ & $-1.25$ \\
2025, reliability reference, 300\,K & $1.9\times10^{13}$ & $2.9\times10^{-19}$ & $6.5\times10^{31}$ & $0.73$ & $-0.40$ & $0.32$ \\
2025, thermal floor, 300\,K & $1.9\times10^{13}$ & $2.9\times10^{-21}$ & $6.6\times10^{33}$ & $0.73$ & $-0.20$ & $0.52$ \\
\multicolumn{7}{l}{\textit{Hypothetical civilizations}} \\
Type I, thermal floor, 2.725\,K (supremum) & $10^{16}$ & $2.6\times10^{-23}$ & $3.8\times10^{38}$ & $1$ & $0$ & $1$ \\
Type II, nucleon confinement at 1\,fm (model value) & $10^{26}$ & $3.3\times10^{-12}$ & $3.0\times10^{37}$ & $2$ & $-1.11$ & $0.89$ \\
Type II, Matrioshka brain, per processor operation & $10^{26}$ & $10^{-16}$ & $10^{42}$ & $2$ & $-0.66$ & $1.34$ \\
Type II, electron confinement at 0.1\,nm (model value) & $10^{26}$ & $6.1\times10^{-19}$ & $1.6\times10^{44}$ & $2$ & $-0.44$ & $1.56$ \\
Type II, thermal floor, 300\,K & $10^{26}$ & $2.9\times10^{-21}$ & $3.5\times10^{46}$ & $2$ & $-0.20$ & $1.80$ \\
Type II, programmable quantum field, 2.725\,K (supremum) & $10^{26}$ & $2.6\times10^{-23}$ & $3.8\times10^{48}$ & $2$ & $0$ & $2$ \\
Type III, thermal floor, 2.725\,K (supremum) & $10^{36}$ & $2.6\times10^{-23}$ & $3.8\times10^{58}$ & $3$ & $0$ & $3$ \\
Ultimate technology, $c^5/4G$, black-hole sink (corner of the plane) & $9.1\times10^{51}$ & $3.6\times10^{-53}$ & $2.5\times10^{104}$ & $4.60$ & $2.99$ & $7.58$ \\
\end{tabular}
\end{ruledtabular}
\end{table*}

\subsection{Civilizations on the scale}
\label{sec:civ}

Table~\ref{tab:civ} places humanity's operations of each class since 1\,CE, a projected fault-tolerant quantum computer, hypothetical Kardashev types and Bradbury's Matrioshka brain on the scale. For humanity's processing in 2025, $K = 0.73$, $B_{\mathrm{floor}} = 0$ and $\mu = B = -0.78$ per switching event at 0.65\,fJ, or 1.5\,fJ of primary energy, which give $\Xi = -0.05$ and $\Omega = 1.3\times10^{28}$ switching events per second. Of $\mu$, $-0.204$ comes from rejecting heat at 300\,K and $-0.572$ from the distance of the gate from the thermal floor at 300\,K. At the device $\mu = -0.74$. Over the published energies per switching event, from 32\,aJ to 3\,fJ, $\mu$ runs from $-0.64$ to $-0.84$. Humanity's electron devices, at 10 to 20\,nm, have the thermal floor at 300\,K to within 0.01 in $B$. The Type II rows on the thermal floor correspond to the rates Wright~\citep{wright2020dyson} computed for a Dyson sphere, and Wright~\citep{wright2023thermodynamics} found little advantage at that floor in nesting shells as a Matrioshka brain does.

\begin{figure*}[p]
\centering
\includegraphics[width=\textwidth]{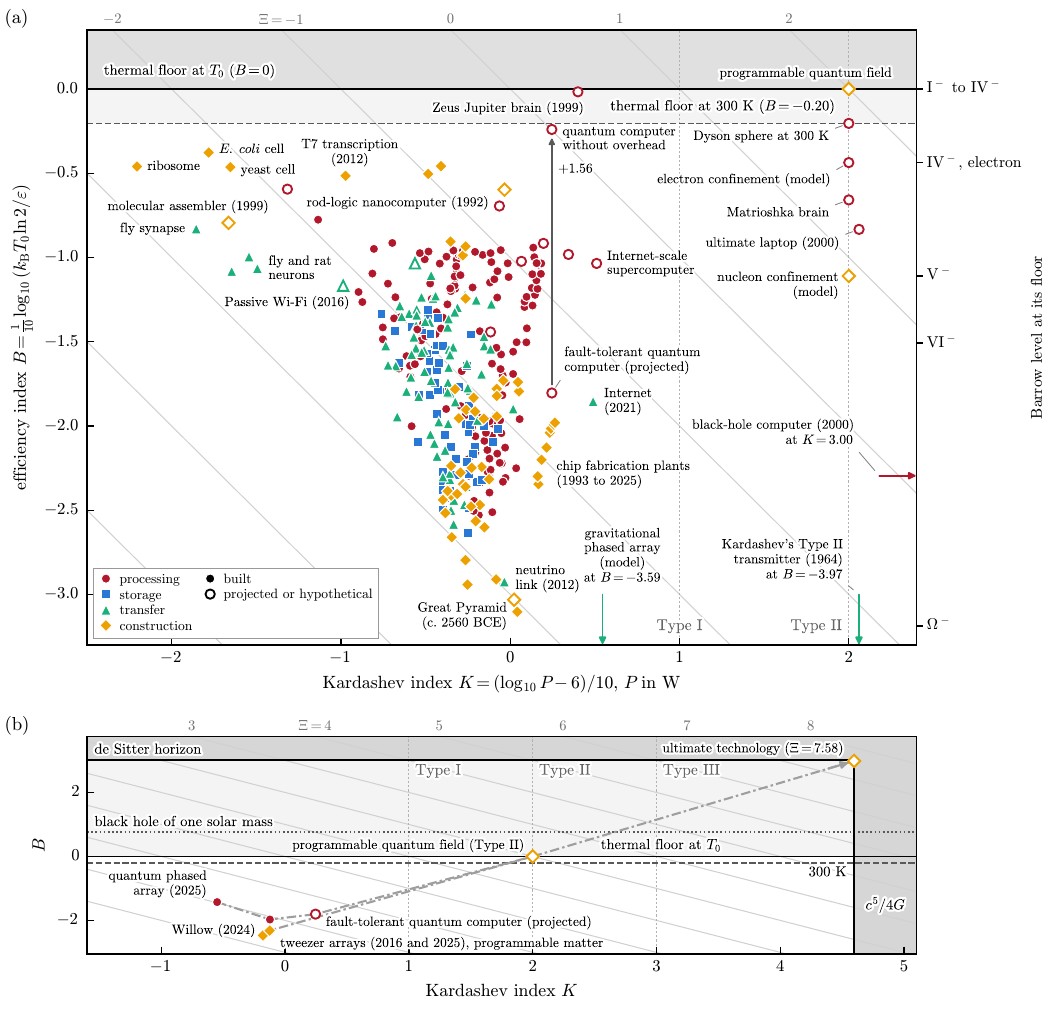}
\caption{Machines, technologies and organisms on the $(K, B)$ plane. (a) All 342 points of Table~\suppref{tab:systems} of the electronic supplementary material and of the Type~II rows of Table~\ref{tab:civ}, each at its own power, labelled outside the windows of Figs.~\suppref{fig:systemsb} to \suppref{fig:systems3} and at a few points within them. Grey diagonals are lines of constant $\Xi = K + B$, dotted lines mark Types I and II, and arrows at the frame point to systems beyond it. The solid and dashed lines are the thermal floors at $T_0$ ($B = 0$) and 300\,K ($B = -0.20$), and the darker band above $B = 0$ is closed at the present epoch to heat radiated into the microwave background. Colour and marker give the class of operation (Section~\ref{sec:budget}), with processing counted per bit of a result. Markers are filled where built and open where projected or hypothetical. The grey arrow is the gain of a quantum computer without overhead over a fault-tolerant machine of the same power, $\Delta\Xi = 1.56$. The right-hand axis marks each level at the $B_{\mathrm{floor}}$ of its example operation (Table~\ref{tab:levels}). (b) The conjectured largest power of one system and the coldest sink (Table~\ref{tab:known}), with grey regions beyond them. The floor of a black hole of one solar mass is dotted, and Types I, II and III are marked. The dash-dotted line joins, in order of power, systems that act on one quantum system at a time, namely the quantum phased array (built), Willow (built), a projected fault-tolerant quantum computer and the programmable quantum field at Type~II (a supremum), and ends at the ultimate technology in the corner of the plane (Section~\ref{sec:eta}). A second dash-dotted line joins to the same field the tweezer arrays (built) that place atoms one by one, the writing operation of programmable matter \citep{gurses2026information}. The dash-dotted lines are guides to the eye.}
\label{fig:systems}
\end{figure*}

Humanity's rows are capacities for its whole power spent on one class of operation. A machine can instead be placed at its own power, and then $\Xi$ counts the operations it performs. Figure~\ref{fig:systems} places in this way 337 machines, technologies and organisms of the four classes, from the builders of the Great Pyramid to machines planned for about 2030, and the five Type~II rows of Table~\ref{tab:civ}. Figures~\suppref{fig:systemsb} to \suppref{fig:systems3} of the electronic supplementary material enlarge windows of Fig.~\ref{fig:systems}, and Table~\suppref{tab:systems} gives values and sources. A single logic gate switching at 3\,GHz has $B = -0.78$ and performs $3\times10^{9}$ operations per second, so $\Xi = -1.91$. El~Capitan, including memory and communication and counted per bit of a 64-bit result, has $B = -1.04$ and, at 68\,MW of primary energy, completes $1.2\times10^{20}$ operations per second, so $\Xi = -0.85$. A system cannot exceed the efficiency of its own switches counted at the same boundary, so its $B$ is at most theirs. From ENIAC, at its own 0.53\,MW, to El~Capitan, $\Xi$ rose by 1.48, of which 0.21 came from the power of the machine and the rest from its energy per operation. A DGX H200 server for artificial intelligence (AI) draws up to 10.2\,kW \citep{nvidia2024dgx} for eight Hopper processors, each performing about $10^{15}$ dense 16-bit floating-point operations per second. Its $B$ of $-0.98$ exceeds that of El~Capitan, but at its lower power its $\Xi$ is $-1.15$. The largest $\Xi$ we found for a built machine, $-0.81$, belongs to a pod of 9216 Ironwood processors, and a campus of $4\times10^{5}$ GB200 processors planned at 1.2\,GW would reach $-0.64$. Run as one Internet-scale supercomputer at the energy per operation of El~Capitan, the data centres of 2025 would reach $K = 0.51$ and $\Xi = -0.52$. The Internet, whose networks drew about 34\,GW of electricity in 2021, has the largest $K$ of the built systems of Table~\suppref{tab:systems}, 0.49, but at $B = -1.85$ per bit delivered its $\Xi$ is $-1.36$. Across the real systems of the four classes the energy per control operation spans 26.5 orders of magnitude, from ${9.8\times10^{-19}}$\,J per bit specified in the protein synthesis of the human body to ${2.8\times10^{8}}$\,J per block placed in the Great Pyramid of Giza. In every family joined by a line in Figs.~\suppref{fig:systemsb} to \suppref{fig:systems3} except the solid-state drives, the optical modules and the printing presses, $B$ rose more than $K$. The leaders of the TOP500 list gained 0.54 in $B$ and 0.24 in $K$ from the CM-5 of 1993 to El~Capitan, and Intel's processors 0.53 in $B$ at the chip from the 4004 to the Core~i9-12900K. Hard disks gained 0.71 in $B$ from the RAMAC of 1956 to a drive of 2023 that draws a three-hundredth of the RAMAC's power, and chip fabrication plants gained 0.37 in $B$ per transistor made and 0.10 in $K$ from 1993 to 2025.

A fabrication plant spends about 5.0\,GJ of electricity \citep{boakes2023cradle}, or 11.6\,GJ of primary energy, on one 300\,mm wafer at the 5\,nm (N5) node. We charge this energy to the $6.9\times10^{12}$ transistors a full wafer holds at the density of the NVIDIA H100 processor, $8.0\times10^{10}$ transistors on 814\,mm$^2$ \citep{andersch2022hopper}. Counting the full wafer, without edge losses or yield, understates the energy per transistor by a small factor. The number of specified bits that set each transistor far exceeds that factor, so $1.7\times10^{-3}$\,J of primary energy per transistor is an upper bound on the energy per specified bit and $-1.98$ is a lower bound on $B$. Fabrication remains costlier per bit than logic unless each transistor is set by more than $10^{12}$ bits. A plant that processes $10^{5}$ wafers per month, a rate that the largest plants exceed \citep{tsmc2025gigafab}, draws 0.44\,GW of primary energy and, at its own power, has $\Xi = -1.72$ per transistor.

Rejecting heat at low temperature requires large radiators, since a sphere of radius $R$ at temperature $T$ rejects a net power $P = 4\pi R^2\sigma(T^4 - T_0^4)$ into the present sky. Wright~\citep{wright2020dyson} uses the same form with surroundings at about 4\,K. For $10^{26}$\,W the radius is $1.3\times10^{11}$\,m at 300\,K and $2.3\times10^{15}$\,m at 3\,K, where the thermal floor has $B = -0.004$. For a Type III civilization it is $1.3\times10^{16}$\,m at 300\,K and $2.3\times10^{20}$\,m at 3\,K. At $T_0$ the radius diverges, so $B = 0$ is a supremum. Wright~\citep{wright2020dyson} notes that ``temperatures near $T_{\mathrm{min}}$ would appear to be strongly disfavored on practical grounds, unless their only purpose is to be cold''. At a fixed radiator area the temperature scales as $P^{1/4}$ once $P$ exceeds the absorbed starlight, so the thermal floor has $B = \mathrm{const} - K/4$, which couples $K$ and $\mu$ at any level. At the Earth's distance from the Sun, a radiator of the Earth's area that also rejects the $1.2\times10^{17}$\,W of sunlight it absorbs runs at 260\,K for $10^{16}$\,W, where sunlight dominates, and at 769\,K for $10^{19}$\,W, placing the thermal floor at $B = -0.20$ and $-0.25$.

\subsection{The trajectory since the first writing}

\begin{figure*}[p]
\centering
\includegraphics[width=135mm]{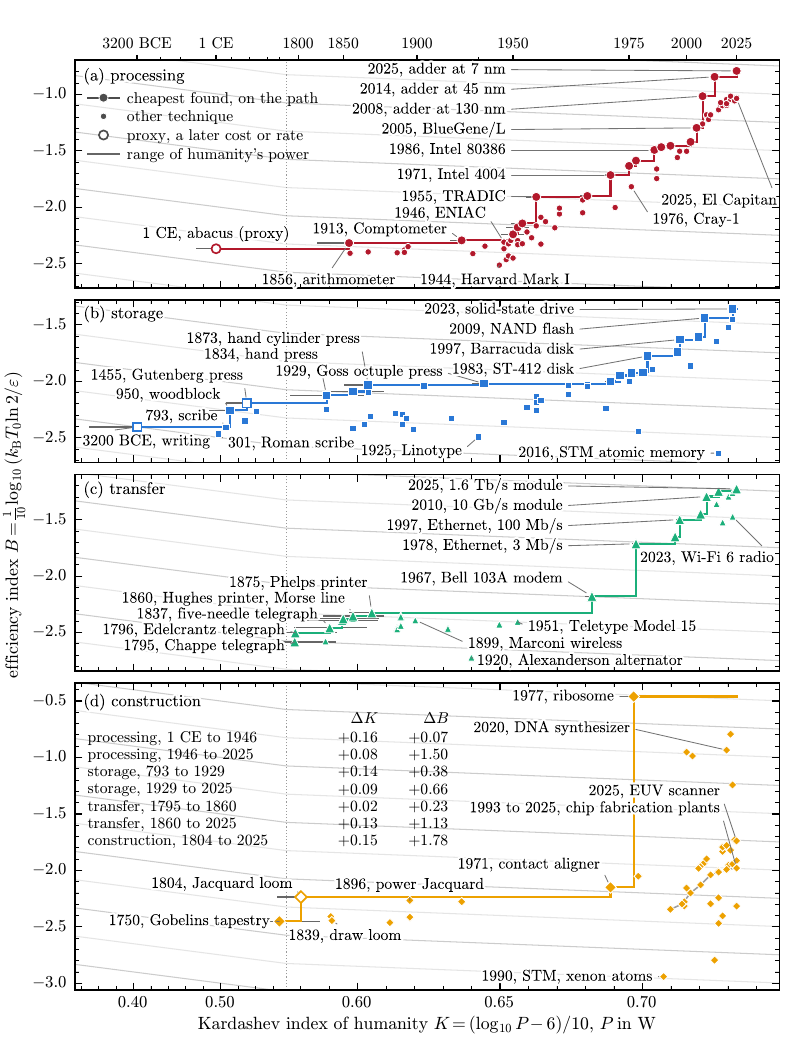}
\caption{Humanity on the $(K, B)$ plane from the first writing to 2025, for (a) processing, (b) storage, (c) transfer and (d) construction. Each point is humanity at its power in the year of a technique, including food and feed, and at the primary energy per operation of that technique. Operations are counted per bit of a result in (a), per bit stored in (b), per bit sent in (c) and per specified bit, transistor, voxel or atom in (d). Large markers joined by a staircase are the cheapest technique found at or before each year, and small markers are other techniques. Open markers are proxies, namely writing on clay in 3200\,BCE at a Carolingian scribe's cost, the abacus of 1\,CE at a 1946 champion's rate, Gutenberg's press of 1455 at the rate set for the wooden press in 1573 and the Jacquard loom of 1804 at a loom speed of 1910. Grey bars give the range of $K$ within the bounds of humanity's energy (electronic supplementary material, Table~\suppref{tab:energy}). Right of the dotted line at $K = 0.575$ the scale of $K$ is 3.3 times larger and the diagonals of constant $\Xi$ bend. The top axis gives the year at which humanity reached each $K$. Sources are in Table~\suppref{tab:history}.}
\label{fig:trajectory}
\end{figure*}

We follow humanity's capacity in each class back to the first writing (Fig.~\ref{fig:trajectory}), taking its power to be its use of energy including food and feed, since before 1950 most operations were powered by muscle. This power grew from $1.1\times10^{10}$\,W in 3200\,BCE to $8.8\times10^{10}$\,W in 1\,CE, $1.6\times10^{12}$\,W in 1900 and $2.1\times10^{13}$\,W in 2025, or from $K = 0.40$ to 0.49, 0.62 and 0.73 \citep{kleingoldewijk2023hyde,owid2024population,malanima2014history,malanima2022world,smil2017civilization}. The bounds of this energy that we derive from published estimates (electronic supplementary material, Table~\suppref{tab:energy}) put $K$ at 0.35 to 0.42 in 3200\,BCE, 0.47 to 0.52 in 1\,CE and 0.616 to 0.628 in 1900. Including food and feed raises $K$ by 0.004 in 1950 and 0.005 in 2025. For each class and year we take the cheapest technique found at or before that year among the 216 of Table~\suppref{tab:history}, which gives their sources. We count transfer per bit sent over a link and charge the operators of manual techniques their metabolic power \citep{herrmann2024compendium}.

An operation of processing cost 12\,J on an abacus at the rate of a 1946 champion, 3.8\,J on the arithmometer of 1856, 2.2\,J on a key-driven Comptometer of 1913 at its maker's timing and 3.2\,J of primary energy on ENIAC. The two calculators are counted per addition within a multiplication, and the abacus per number added. From 1950 the cheapest technique was, in order, Whirlwind I, SWAC, the IAS machine, the TRADIC transistor computer of 1955, the Apollo Guidance Computer of 1966, microprocessors from the Intel 4004 of 1971 to the Pentium of 1993, the Earth Simulator, BlueGene/L and adders at 130, 45 and 7\,nm. With the abacus as the technique of 1\,CE, $\Xi$ for processing rose by 0.23 from 1\,CE to 1946, of which 0.16 came from $K$, and by 1.58 from 1946, when the Comptometer was still the cheapest, to 2025, of which 0.08 came from $K$. Within the bounds of humanity's energy the first rise in $K$ is 0.12 to 0.18.

From 29\,J per bit for a Carolingian scribe in 793, printing lowered the cost to 0.97\,J with woodblocks in China by 950, 0.22\,J with Gutenberg's press at the rate set for the wooden press in 1573 and 4.2\,mJ of primary energy with the electric octuple rotary presses of 1929, counting their electricity at the factor of 1950. The hand press of 1834, a steam press of 1860 and a hand-cranked cylinder press of 1873 fill in the path of Fig.~\ref{fig:trajectory}, the last cheaper per bit than the steam presses of its time. Printing remained the cheapest storage we found until the IBM 3330 disk of 1971. From 793 to 1929 printing raised $B$ for storage by 0.38 while $K$ rose by 0.14, or 0.11 to 0.17 within the bounds of humanity's energy, and from 1929 to 2025 electronic storage raised $B$ by 0.66 more while $K$ rose by 0.09.

A bit sent by Chappe's optical telegraph in 1795 cost 1.7\,kJ, and one sent by Edelcrantz's shutter telegraph in 1796 cost 0.27\,kJ. Needle telegraphs lowered the cost to 98\,J in 1837 and 23\,J in 1849, and printing telegraphs to 8.8\,J with Hughes's instrument of 1860 and 4.6\,J with Phelps's of 1875. A Morse line of 1860 spent 10\,J per bit, almost all of it the metabolic energy of its two operators. Early radio cost 20\,J to $4\times10^{4}$\,J per bit, and no technique we found sent a bit more cheaply than the telegraphs until the Bell 103A modem of 1967. An optical module now sends one for 49\,pJ of primary energy. From 1795 to 1860 the telegraphs raised $B$ for transfer by 0.23, or by 0.21 with every telegraph counted per character, while $K$ rose by 0.02, by at most 0.05 within the bounds of humanity's energy. From 1860 to 2025 $B$ rose by 1.13 more and $K$ by 0.13.

Weaving a Gobelins tapestry in 1750 cost 87\,J per bit of the colour chosen for each cell, a warp thread set by a Jacquard loom in 1804 cost 0.61\,J at the loom speed of 1910, and a transistor exposed at one mask level by a manual contact aligner in 1971 cost 84\,mJ. Enzymes directed by designed genes have set bits for about $10^{-18}$\,J since 1977. From 1804 to 2025 $B$ for construction rose by 1.78 and $K$ by 0.15. Since the first documented cost, $B$ supplied 92\% of the gain in $\Xi$ for processing (from 1856), 82\% for storage (from 793) and 90\% for transfer (from 1795), and since the Jacquard loom of 1804, whose rate is a proxy, it supplied 92\% for construction. Per transistor made by a fabrication plant instead of per bit set by a ribosome, $B$ supplied 63\% of the gain in construction since 1804.

For processing since 1950 we divide the change in $\Omega$ between its numerator and its denominator. World primary energy was 100.7\,EJ in 1950, or $3.2\times10^{12}$\,W and $K = 0.65$ \citep{smil2017energy,owid2026primaryenergy}. This total includes 27\,EJ of traditional biomass, which the 2025 total excludes. Between ENIAC and El~Capitan, each counted for the whole machine, the primary energy per operation fell by 12.7 orders of magnitude and $P$ rose by 0.8 orders, so $\Xi$ for processing rose by 1.35, of which 0.08, or 6\%, came from $K$. Counting traditional biomass consistently in both years gives $\Delta K$ of 0.08 to 0.09. At the machines' own power the rise is 1.48 (Section~\ref{sec:civ}). The costs of the two machines refer to bits of different results. At a single granularity the rise in $\Xi$ is 1.59 for addition, from ENIAC to a 32-bit addition at 7\,nm, and 1.27 for a switching event, from a device of 1950, by its date a vacuum tube, at $3.4\times10^{-4}$\,J in the compilation of Keyes published by Landauer \citep{landauer1988dissipation,lee2017cmosdata}, to the IRDS gate. The share of the denominator is 94\% to 95\% at every granularity. Koomey \textit{et al.}~\citep{koomey2011implications} found that computations per joule doubled every 1.57 years from 1946 to 2009, and Naffziger \& Koomey~\citep{naffziger2016energy} found that the doubling time has lengthened to about 2.6 years since 2000.

Humanity held III-minus before 1950, reached II-minus in the 1970s and IV-minus in the laboratory in 1990, and all three levels have the thermal floor. A vacuum tube and a transistor are electron devices with that floor, so the fall of 12.7 orders in $\eps$ is a narrowing of the maturity gap by 1.27. About 0.2 of these orders came from more efficient generation of electricity, and the rest from the devices. Part of this narrowing is the inward trend Barrow described, the miniaturization of devices within the levels above the atom. We estimate it from the switching energy $CV^2$ of a gate, whose capacitance falls in proportion when all dimensions are reduced by a common factor \citep{dennard1974design}. From a feature size of 10\,$\mu$m at 5 to 15\,V to about 20\,nm at the 0.7\,V of the IRDS gate, $CV^2$ fell by 4.4 to 5.4 orders of magnitude, 2.7 of them from the capacitance. Per switching event at the device, from a vacuum tube to the IRDS gate, the energy fell by 11.7 orders. Miniaturization accounts for 4.4 to 5.4 of them, an upper estimate since oxide thickness and supply voltage stopped scaling before the feature size did. The step from the tube to a 10\,$\mu$m transistor, which accounts for the remaining 6.3 to 7.3, also involved shrinking but changed the physics of the device, and the estimate does not separate the two.

Per specified bit, humanity's cheapest operation since the late 1970s has been enzymatic. Charged at the primary-energy boundary for the glucose behind their ATP and monomers, the ribosomes and polymerases humanity directs lie at $B = -0.46$ and $-0.52$, against $\mu = -0.78$ for its logic (Section~\ref{sec:civ}) and $B = -1.98$ for its fabrication. At the device, without glucose or monomers, the enzymes are at $-0.36$. Since copying a template has no floor per bit (Section~\ref{sec:levels}), their distances to the thermal floor are not maturity gaps.

\subsection{The remaining maturity gap}
\label{sec:gap}

At 300\,K the IRDS gate lies 5.7 orders of magnitude above the thermal floor and 3.7 above the reliability reference, or 5.4 and 3.3 at the device, and the rest of its 7.8 orders above the floor at $T_0$ (Section~\ref{sec:civ}) comes from rejecting heat at 300\,K instead of at $T_0$. Humanity's power is 2.7 orders below Sagan's Type I and exceeds Kardashev's Type I, his estimate of humanity's consumption in 1964, by 0.7 orders, so it is below Type I only in Sagan's convention, which we use throughout. Closing the maturity gap at the present power would raise $\Xi$ for processing by 0.78, although its last 0.20 is a supremum that needs an ever larger radiator. Bringing gates to the thermal floor at 300\,K would raise $\Xi$ by 0.57, and to the reliability reference by 0.37, while reaching Type I with present gates would raise it by 0.27. The comparison with Type I holds for every published switching energy of complementary metal-oxide-semiconductor (CMOS) logic, since even a bare transistor at 32\,aJ lies 6.4 orders above the thermal floor at $T_0$ in primary energy, and 4.4 above that at 300\,K. Against the reliability reference, the comparison holds for gate energies above 0.07\,fJ at the device. The IRDS roadmap projects 0.40\,fJ per switching event from 2028 in its 2023 edition and 0.33\,fJ from 2033 in its 2024 edition, at 0.60\,V \citep{irds2023moremoore,irds2024moremoore}. Both projections are 5.4 to 5.5 orders above the thermal floor at 300\,K in primary energy and 5.1 at the device.

Ayala \textit{et al.}~\citep{ayala2021mana} reported 1.4\,zJ per Josephson junction for superconducting adiabatic logic at 4.2\,K, and Takeuchi \textit{et al.}~\citep{takeuchi2013measurement} measured about 10\,zJ per gate bit. Cooling costs 70\,W of work per watt removed at the Carnot limit, about 230\,W/W in large helium plants and up to $10^{4}$\,W/W in small cryocoolers \citep{radebaugh2004refrigeration}. At the wall these costs become 0.1 to 100\,aJ and lie 1.5 to 4.5 orders above the thermal floor at 300\,K, or 1.9 to 4.9 orders in primary energy, against 5.7 for CMOS. The comparison with Type I above refers to CMOS, which performs almost all of humanity's logic.

A fault-tolerant universal quantum computer computes with unitary gates, which are reversible and not counted. It removes the entropy that noise adds to its qubits by measuring and resetting its measure qubits in every cycle of error correction. Each reset is a control operation, and the classical syndrome bits the cycle produces are not counted. The 105-qubit surface-code processor Willow suppressed errors below threshold with 48 measure qubits and a cycle of 1.1\,$\mu$s \citep{acharya2025quantum}. At the 26\,kW that Arute \textit{et al.}~\citep[Supplementary Information]{arute2019quantum} estimated for the apparatus of its predecessor, Willow spends $6.0\times10^{-4}$\,J per reset at the wall, or $1.4\times10^{-3}$\,J of primary energy and $B = -1.97$. Among comparable systems, IQM states 26\,kW for its 54-qubit Radiance \citep{iqm2026radiance} and Carrasco-Codina \textit{et al.}~\citep{carrascocodina2026energy} measured 21.4\,kW for the 32-qubit QMIO. The quantum annealer Advantage2 reads its 4575 qubits and returns them to their initial state in every sample, which takes about 180\,$\mu$s with the default anneal, the longest readout and the delay between samples. At the 12.5\,kW its maker states, it spends $1.1\times10^{-3}$\,J of primary energy per reset, so $B = -1.96$ \citep{dwave2025advantage2,dwave2026qpuprops}. A machine of $2\times10^{7}$ physical qubits with a cycle of 1\,$\mu$s could factor a 2048-bit number in 8 hours \citep{gidney2021how} and would draw about 125\,MW \citep{parker2023estimating}. If half of its qubits are reset in each cycle, it spends $1.25\times10^{-5}$\,J per reset at the wall, or $2.9\times10^{-5}$\,J of primary energy and $B = -1.80$ (Table~\ref{tab:civ}). Its qubits are held near 10\,mK, but its heat is finally rejected near 300\,K, so its primary energy per reset lies 16 orders of magnitude above the thermal floor at 300\,K and 18 above the floor at $T_0$. A quantum computer without overhead would spend $\kB T\ln 2$ of electricity per counted reset at 300\,K, or 2.3 times that in primary energy. Its least heat per reset is lower. Given the outcome of the previous cycle, a measure qubit whose outcome changes with probability $p$ holds only $H_2(p)$ bits before reset, 0.43 bits at the detection probability of 8.7\% measured on Willow at distance 7 \citep{acharya2025quantum}. With the same power a quantum computer without overhead would reset $4.4\times10^{28}$ qubits per second at $B = -0.24$, about $10^{23}$ qubits at the same cycle, and at its own power it would have $\Xi = 0.01$, 0.86 above El~Capitan (Fig.~\ref{fig:systems}).

Once an operation that resets a thermally random degree of freedom reaches the thermal floor at its rejection temperature, $\Xi$ can grow only with $K$, with a colder sink or with a new sink for entropy (Eq.~\eqref{eq:OmegaS}). Humanity's $K$ has grown by 0.001 per year since 1950, and colder radiators can add at most 0.20 at the present epoch to a civilization that rejects its heat at 300\,K. Reversible logic can operate below the floor at the cost of area and latency \citep{bennett1973logical,athas1994low}. For a civilization that computes reversibly, $\Omega$ bounds the rate at which it writes results into storage and resets thermally random degrees of freedom of matter, while its logic has no such bound (Section~\ref{sec:levels}).

\subsection{Power from the levels below the atom}
\label{sec:eta}

A Type II civilization spending its power on nucleon-scale operations at 21\,MeV each would perform $3\times10^{37}$ per second, 11 orders fewer than on the thermal floor at $T_0$. Barrow described the lower levels as ``engineering the nucleons'' and creating ``organized complexity among populations of elementary particles''. Since the floor is the same at every level (Section~\ref{sec:levels}), the lower levels raise $\Omega$ only through $P$, of which fusion and fission are sources. {\'C}irkovi{\'c}~\citep{cirkovic2015kardashev} noted that a V-minus civilization could draw power more efficiently by running fusion than by harvesting a star's output with a Dyson shell. The power drawn from fuel is
\begin{equation}
P = \eta\,\dot M_{\mathrm{f}}\,c^{2},
\label{eq:eta}
\end{equation}
where $\dot M_{\mathrm{f}}$ is the rest mass of fuel and oxidizer consumed per second and $\eta$ the fraction of its rest energy the process releases (Table~\ref{tab:eta}). Matter accreting onto a black hole through a thin disc releases the binding energy of the innermost stable circular orbit \citep{bardeen1972rotating}, 5.7\% for a hole without spin and 32\% at the spin of 0.998 that accretion can give it \citep{thorne1974disk}. The rotational energy of a maximally spinning hole is 29\% of its mass \citep{christodoulou1970reversible}. Black-hole evaporation \citep{hawking1975particle} and annihilation release the whole rest energy. With $K_{\mathrm{f}} = (\log_{10}\dot M_{\mathrm{f}}c^2 - 6)/10$ and $K_\eta = \tfrac{1}{10}\log_{10}\eta$, the identity $K = K_{\mathrm{f}} + K_\eta$ gives
\begin{equation}
\Xi = K_{\mathrm{f}} + K_\eta + B_{\mathrm{floor}} + \mu .
\label{eq:Xi4}
\end{equation}
The levels enter $\Xi$ twice, in $K_\eta$ through the reaction that makes the power and in $B_{\mathrm{floor}}$ through the level at which operations act. Unlike $B_{\mathrm{floor}}$, $K_\eta$ does not require holding a level in our sense, since a reactor specifies no state for each nucleus. At a fixed rate of consumption, moving from burning hydrogen in oxygen to fusing hydrogen adds 0.76 to $K$, and complete conversion adds 0.98, nearly one Kardashev type. About 86\% of humanity's total energy supply comes from fossil fuels \citep{ei2026statistical}, whose combustion consumes about $1.7\times10^{6}$\,kg of fuel and oxygen per second, so $K_{\mathrm{f}} = 1.72$ and $K_\eta = -1.00$. The Sun burns about $6\times10^{11}$\,kg of hydrogen per second, so $K_{\mathrm{f}} = 2.27$ and $K_\eta = -0.21$.

Local conservation of energy bounds $\eta$ by 1, and $K_\eta$ by 0, when $\dot M_{\mathrm{f}}$ includes every mass consumed, such as the mass a black hole loses as its spin is extracted. Tchekhovskoy \textit{et al.}~\citep{tchekhovskoy2011efficient} found in simulations of magnetically arrested accretion that a spinning hole can return about 140\% relative to the accreted matter alone. Energy is not conserved globally in an expanding universe, and Harrison~\citep{harrison1995mining} argued that work can in principle be extracted from the expansion, so the bound applies to power drawn from rest mass. Since $K_\eta \le 0$, and $B_{\mathrm{floor}} \le 0$ and $\mu \le 0$ when entropy leaves as heat radiated into the present sky, Eq.~\eqref{eq:Xi4} gives $\Xi \le K_{\mathrm{f}}$, which bounds the operations a civilization can perform at the thermal floor by the rest energy of its fuel. A programmable quantum field, a name we introduce for the limiting constructor, would set the state of every mode of every field within its reach, and so perform operations of every class. One that released the whole rest energy of its fuel ($K_\eta = 0$) and set every degree of freedom arbitrarily close to the thermal floor at $T_0$ without overhead ($B_{\mathrm{floor}} = 0$ and $\mu \to 0$) would approach $\Xi = K_{\mathrm{f}}$, the bound set by its rate of fuel consumption. At humanity's present flow of fuel this bound is $\Xi = 1.72$, and at the power of a Type II civilization it is $\Xi = 2$, the supremum listed for the programmable quantum field in Table~\ref{tab:civ} (Fig.~\ref{fig:systems}).

A built proxy of a programmable quantum field is a quantum phased array, a coherent array of antenna elements that transmits or receives quantum fields such as non-classical states of light \citep{gurses2025onchip,gurses2026information}. The first realization receives squeezed light over free space through 32 antennas, applies a programmed phase and weight to the mode of each with 32 thermo-optic phase shifters, combines the modes and reads a quadrature in each of 32 channels by homodyne detection over a bandwidth of 10.4\,MHz \citep{gurses2025onchip}. The squeezed light is made off the chip, so the device reads and manipulates the field and does not yet set its state. We do not count the measurement (Section~\ref{sec:known}). Each sample overwrites a register, and by the rule of Section~\ref{sec:budget} we count one bit per sample. Two samples per second per hertz of bandwidth in each channel give $6.7\times10^{8}$ operations per second, a lower bound since the digitizers of the experiment resolve 14 bits. We place the array in processing, charge it at the chip and show it with the points of Fig.~\suppref{fig:systems2}(a). At the 1.4\,W its thermo-optic phase shifters draw at full drive, it has $K = -0.55$ and $B = -1.43$ (Table~\suppref{tab:systems}). Figure~\ref{fig:systems}(b) orders it with Willow, the fault-tolerant quantum computer of Section~\ref{sec:gap} and the programmable quantum field, as systems that act on one quantum system at a time.

Lloyd's ultimate laptop, 1\,kg in 1\,L, must radiate ${4.0\times10^{26}}$\,W at ${5.9\times10^{8}}$\,K to reject erroneous bits at the rate its surface allows \citep{lloyd2000ultimate}. Counted by the bits it radiates, since its reversible operations are not counted, it lies at $K = 2.06$ on the thermal floor at its own temperature. Lloyd~\citep{lloyd2000ultimate} and Lloyd \& Ng~\citep{lloyd2004black} took a black hole of 1\,kg as a computer whose input falls in and whose result is read from its Hawking radiation. The hole empties its $3.8\times10^{16}$ bits in about $10^{-19}$\,s at its own temperature of $1.2\times10^{23}$\,K, so $K = 3.00$ and $B = -2.30$. A large black hole is instead a cold sink, since heat dropped into it raises its entropy by the heat divided by its temperature \citep{bekenstein1974generalized,hawking1975particle}. An uncharged, non-rotating black hole in de Sitter space is coldest at the largest mass possible for it (Section~\ref{sec:known}). A system of the conjectured largest power, $c^5/4G$, that drops its entropy into such a black hole marks the corner of the $(K, B)$ plane at $K = 4.60$ and $B = 2.99$, or $\Xi = 7.58$. We call it the ultimate technology (Table~\ref{tab:civ}, Fig.~\ref{fig:systems}(b)). Both coordinates are limits of Section~\ref{sec:known}, one conjectured and one a supremum for a sink that is a thermal bath or an uncharged, non-rotating black hole, and no system reaches the corner. As a black hole approaches the largest mass, the entropy it can still absorb and the energy left in its causal patch decrease, and at $c^5/4G$ the energy of a patch is spent in ${4\times10^{17}}$\,s, under one de Sitter time.

Toffoli \& Margolus~\citep{toffoli1991programmable} defined programmable matter as a computing medium whose physical properties are set by computation. Goldstein \& Mowry~\citep{goldstein2004claytronics} pursued it with robots of a millimetre, and at the scale of the atom Barredo \textit{et al.}~\citep{barredo2016atom} and Endres \textit{et al.}~\citep{endres2016atom} rearranged neutral atoms one by one in optical tweezers, the tweezer arrays of Fig.~\ref{fig:systems}(b) \citep{gurses2026information}. A gravitational phased array, an array of coherently driven sources of gravitational waves \citep{gurses2026information,romero1981generation}, would act on the gravitational field. Barrow placed the structure of space and time at $\Omega$-minus, where we identify no control operation (Table~\ref{tab:levels}). We place on the scale the representative element of Gurses~\citep{gurses2026information}, a point mass of 1\,kg oscillating at 1\,GHz with an amplitude of 1\,nm about the centre of the body that holds it. By the quadrupole formula it radiates $(16/15)\,G m^2 A^4\omega^6/c^5 = 1.8\times10^{-30}$\,W at 2\,GHz, twice the frequency of its motion, so each graviton carries $1.3\times10^{-24}$\,J. The element stores 20\,J of mechanical energy and at a quality factor of $10^{4}$ draws $1.2\times10^{7}$\,W of drive, so each graviton costs $9.1\times10^{12}$\,J of drive. Ten thousand such elements would have $K = 0.55$ and $B = -3.59$ (Table~\suppref{tab:systems}). The element is a model, since its peak acceleration of $4\times10^{9}$\,g would stress a solid of its size to about $10^{13}$\,Pa, two orders of magnitude above the ideal strength of any material, and a solid can move at 1\,GHz only as an acoustic resonator, whose quadrupole moment is set by its strain field and its geometry rather than by a rigid displacement. In primary energy the element lies 36 orders of magnitude above the thermal floor at $T_0$, against 7.8 for humanity's logic gates. The drive per graviton depends on the geometry of the source, and a stack of plates vibrating in their thickness mode would lie about 31 orders above the floor (electronic supplementary material, section~\suppref{sec:inputs}). The thermal floor bounds every transfer operation whatever its carrier, so a transmitter that set the state of a gravitational mode would have $B \le 0$ at the present epoch, as a transmitter of photons does. At 2\,GHz the floor is thermal, since a graviton of the array carries a twentieth of $\kB T_0\ln 2$. We know of no process that sets the quantum state of a gravitational mode, and the array above produces a field of about $10^{-7}$ gravitons per mode. On the receiving side, observatories of gravitational waves are arrays that read the gravitational field \citep{gurses2026information}, and those on the ground are quantum-enhanced by squeezed light \citep{tse2019quantum}. Stancil \textit{et al.}~\citep{stancil2012demonstration} sent 0.1 bit per second with a beam of neutrinos from 120\,GeV protons over 1.0\,km, 240\,m of it through earth, at $K = -0.03$ and $B = -2.92$. Tiedau \textit{et al.}~\citep{tiedau2024laser} and Zhang \textit{et al.}~\citep{zhang2024frequency} excited the isomeric state of the thorium-229 nucleus with lasers, though no single nucleus has yet been set to a specified state this way. Bose \textit{et al.}~\citep{bose2017spin} and Marletto \& Vedral~\citep{marletto2017gravitationally} proposed tests that entangle two masses through their gravitational interaction, a reversible operation that sets no degree of freedom of spacetime and so is not a control operation at $\Omega$-minus.

\begin{table}[t]
\caption{Fraction $\eta$ of the rest energy of the fuel released by processes a civilization could use, and $K_\eta = \tfrac{1}{10}\log_{10}\eta$. Combustion is per kilogram of hydrogen and oxygen. Including the energy carried off by neutrinos and gravitons in $\eta$ overstates the power commanded by less than 0.003 in $K$ for the nuclear processes and by 0.03 for proton-antiproton annihilation, in which neutrinos carry about half the energy. Accretion is the binding energy of the innermost stable circular orbit for spin $a$, and the spin energy is per unit mass of the hole. Annihilation is a source only for naturally occurring antimatter. GR denotes general relativity.}
\label{tab:eta}
\begin{ruledtabular}
\begin{tabular}{llll}
Process & Physics & $\eta$ & $K_\eta$ \\
\hline
H$_2$ + O$_2$ combustion & chemistry & $1.8\times10^{-10}$ & $-0.98$ \\
$^{235}$U fission & nuclear & $9.2\times10^{-4}$ & $-0.30$ \\
Deuterium-tritium fusion & nuclear & $3.8\times10^{-3}$ & $-0.24$ \\
H to $^{4}$He fusion & nuclear & $7.1\times10^{-3}$ & $-0.21$ \\
Accretion, $a = 0$ & GR & $0.057$ & $-0.12$ \\
Spin energy, $a \to 1$ & GR & $0.29$ & $-0.05$ \\
Accretion, $a = 0.998$ & GR & $0.32$ & $-0.05$ \\
Evaporation & quantum, GR & 1 & 0 \\
Annihilation & particles & 1 & 0 \\
\end{tabular}
\end{ruledtabular}
\end{table}

\section{Storage and transfer}
\label{sec:companions}

Storage and transfer consist of control operations, which are counted in $\Omega$, but storage also draws on mass and transfer on channels. Mass also limits construction to $\Omega \le \dot M/m_{\mathrm{op}}$, where $\dot M$ is the rate at which matter is processed and $m_{\mathrm{op}}$ the mass per operation. At humanity's power on the reliability reference at 300\,K, one operation per atom of 26\,u, the Earth's mean, would consume $2.8\times10^{6}$\,kg of matter per second.

A passive medium holds a bit behind a barrier for a time set by the barrier's height and spends energy only when the bit is written. The number of bits a civilization can hold is therefore the mass it commits divided by the mass per bit, $N_S = M/m_{\mathrm{bit}}$. The mass per bit is that of a monomer at II-minus, a molecule at III-minus, an atom at IV-minus and a nucleon at V-minus. At one bit per nucleon a kilogram holds $6.0\times10^{26}$ bits, $A$ times more than at one bit per atom of mass number $A$ and, at nuclear density, 15 orders of magnitude more per unit volume. Extrapolating the census of Hilbert \& L{\'o}pez~\citep{hilbert2011world} gives about $10^{23}$ bits held by humanity, and the Earth at one bit per atom would hold $1.4\times10^{50}$. In the form Gray~\citep{gray2020extended} gives for his information scale $K_I$, which we write $\Sigma = (\log_{10}N_S - 6)/10$, these are $\Sigma = 1.7$ and 4.4. Gray's units follow Sagan's lettered information scale \citep{sagan1973cosmic}. A black hole reaches the Bekenstein bound \citep{bekenstein1981universal} and holds $4\pi G M^2/\hbar c\ln 2$ bits \citep{lloyd2000ultimate}, more than one per nucleon above $1.6\times10^{10}$\,kg. A civilization of $10^{16}$\,W on the thermal floor at $T_0$ would set one bit in each atom of the Earth in $3.6\times10^{11}$\,s, less than $10^{-6}$ of a Hubble time.

The floor per received bit is again $\kB T\ln 2$, as Shannon's limit for a channel with thermal noise at $T$ and as the least heat of writing each received bit into matter. Kardashev's $100\,\kB T_N$ per bit is 144 times this. A deployed 800\,Gb/s optical module dissipates up to 17\,W, about 21\,pJ per bit \citep{nvidia2025mms4x00}. A pure-loss channel has no floor, since its energy per bit tends to zero as the number of photons per mode falls. At $10^{-3}$ photons per mode the Holevo capacity is 11.4 bits per photon \citep{holevo1973bounds,giovannetti2004classical}, or 11\,zJ per bit at 1550\,nm, the band of humanity's optical fibres. The microwave background adds thermal noise to every mode of the present sky, so the floor applies again, and a receiver open to the sky needs at least $\kB T_0\ln 2$ per bit \citep{giovannetti2014ultimate}. The capacity of a single channel grows at most as the square root of its power, $\dot I \le \sqrt{\pi P/3\hbar}/\ln 2$ \citep{pendry1983quantum}. A Type II civilization putting $10^{26}$\,W into one channel would reach $1.4\times10^{30}$ bits per second, 18 orders of magnitude below its operation budget on the thermal floor at $T_0$, so it needs many channels to use its power.

\section{Assumptions and scope}
\label{sec:known}

\begin{table*}[t]
\caption{Assumptions on which the limits of the scale rest, the limits they set and their values in this paper.}
\label{tab:known}
\begin{ruledtabular}
\begin{tabular}{>{\raggedright\arraybackslash}p{4.4cm}>{\raggedright\arraybackslash}p{5.0cm}>{\raggedright\arraybackslash}p{6.6cm}}
Rests on & Limit & Value \\
\hline
Second law, including horizons & entropy disposed of per operation, Eq.~\eqref{eq:OmegaS} & $\kB\ln 2$ per bit \\
Coldest accessible sink & $B$ for heat rejected to a sink & $0$ into the present sky, $+3.01$ in de Sitter space at vanishing power and $+2.99$ into the largest uncharged, non-rotating black hole, for a total of at most $\pi\kB/\Lambda\ell_{\mathrm{P}}^2$ \\
Local conservation of energy & $K_\eta$, Eq.~\eqref{eq:Xi4} & $\eta \le 1$, $K_\eta \le 0$ \\
Maximum force, conjectured & power of one system & $K \le 4.60$ \\
Constant $\Lambda > 0$, the generalized second law and the N-bound & irreversible operations in one causal patch & about $5\times10^{122}$ \\
Operation and carrier of each level & model values below the atom & $B_{\mathrm{floor}} = -1.11$ and $-1.51$ \\
General relativity and quantum mechanics extrapolated to $\ell_{\mathrm{P}}$, with no new physics in between & depth of the Barrow scale & $\Omega$-minus at the Planck energy, with $B = -3.19$ for a massless quantum and no operation identified \\
Linear quantum mechanics & rate per unit energy & Margolus-Levitin bound \\
& operations per second of static time in one causal patch of the Nariai mass & $2.3\times10^{103}$, $\Xi \le 7.48$ \\
& capacity of a channel & Holevo and Pendry bounds \\
\end{tabular}
\end{ruledtabular}
\end{table*}

The floors and limits of this paper rest on $\mathcal{L}$, and limits of this kind have moved before by orders of magnitude in both directions. Fusion \citep{bethe1939energy} supplies a star with more than a hundred times the energy of the gravitational contraction once thought to power the Sun. Landauer~\citep{landauer1961irreversibility} restricted the cost of information processing to erasure, and Bennett~\citep{bennett1973logical} showed that logic needs no minimum dissipation. Dyson~\citep{dyson1979time} found that life in an open universe could process an unbounded amount of information with finite energy, while after Riess \textit{et al.}~\citep{riess1998observational} and Perlmutter \textit{et al.}~\citep{perlmutter1999measurements} discovered accelerated expansion, Krauss \& Starkman~\citep{krauss2000life} found the recoverable information finite, ``in the absence of possible exotic and uncertain strong gravitational effects''. Table~\ref{tab:known} lists the assumption on which each limit rests.

The Planck power $c^5/G = 3.6\times10^{52}$\,W gives $K = 4.66$, and the maximum force $c^4/4G$ that Gibbons~\citep{gibbons2002maximum} conjectured in general relativity would bound the power of any one system by $c^5/4G$, or $K = 4.60$ \citep{barrow2015maximum}. Jowsey \& Visser~\citep{jowsey2021counterexamples} proposed counterexamples to the conjecture, and Cardoso \textit{et al.}~\citep{cardoso2018remarks} found luminosities up to $0.2\,c^5/G$ in simulations of Einstein's equations, and arbitrarily large ones for initial data that contain a past horizon. With a positive cosmological constant $\Lambda$, the de Sitter horizon radiates at $T_{\mathrm{dS}} = \hbar H_\Lambda/2\pi\kB = 2.2\times10^{-30}$\,K for an observer at rest at the centre of its patch, with $H_\Lambda = c\sqrt{\Lambda/3}$ \citep{gibbons1977cosmological,planck2020cosmological}. Once the microwave background has cooled below it, the horizon is the coldest thermal bath. An uncharged, non-rotating black hole in de Sitter space is never colder than the cosmological horizon of the same spacetime. Its temperature, measured by the observer who needs no acceleration to stay at rest between the two horizons, falls with increasing mass to $\sqrt{3}\,T_{\mathrm{dS}}$ at the largest mass possible for it, the Nariai mass $M_{\mathrm{N}} = c^3/3\sqrt{3}\,GH_\Lambda = {4.3\times10^{52}}$\,kg, where both horizons have this temperature \citep{bousso1996pair}. A charged black hole can be heavier, approaching $\sqrt{2}\,M_{\mathrm{N}}$, and a charged or rotating one near extremality is colder, since its temperature vanishes at extremality \citep{romans1992supersymmetric,booth1999cosmological}, so the limits on $B$ below refer to thermal baths and to uncharged, non-rotating black holes. The horizon bounds $B$ for heat rejected to a sink by $+3.01$, a supremum approached only at vanishing power. Heat dropped into a black hole near the largest mass has the floor $B = 2.99$ at any power. The entropy the hole can still take vanishes as it approaches that mass, and it takes at most $\pi/\Lambda\ell_{\mathrm{P}}^2 \approx {1\times10^{122}}\,\kB$ over its whole growth, a third of what the horizon holds (Section~\ref{sec:eta}). A civilization confined to one causal patch can dispose of no more entropy than the horizon holds, $3\pi/\Lambda\ell_{\mathrm{P}}^2 \approx 3\times10^{122}\,\kB$, by the N-bound and the generalized second law \citep{bousso2000positive}, so by Eq.~\eqref{eq:OmegaS} it can perform at most about $5\times10^{122}$ irreversible operations in all, whether the entropy goes into heat or into a stock of low-entropy matter. The Margolus-Levitin bound at the Nariai mass, $2M_{\mathrm{N}}c^2/\pi\hbar$, caps the rate of operations of any kind in the patch at ${2.3\times10^{103}}$ per second of static time, or $\Xi = 7.48$. Near the Nariai mass the proper time of the observer who needs no acceleration runs slower than the static time by a factor that vanishes at that mass, so this cap does not bound the corner of the plane. The Barrow scale ends at the Planck length, about 17 orders of magnitude in energy beyond VI-minus, and any physics found in between would add levels.

The measured positions of humanity's devices, the decomposition of the trajectory since 1950 and the estimate for miniaturization are independent of $\mathcal{L}$. For operations that reset thermally random degrees of freedom, the result that descending the levels cannot lower the floor depends only on the second law and the counting of states, which apply at every level. It holds after any discovery that changes the floor of every level together, such as a colder sink, and would fail only if a sink or a relaxation of the second law were open to some level and closed to the levels above it, or if new physics added a cost absent at the lower levels. The entries below the atom depend most on the operation and carrier assigned to each level. A $^{208}$Pb nucleus confined to 1\,fm has $B_{\mathrm{floor}} = -0.88$, against $-1.11$ for a nucleon.

The scale does not measure the value, design or function of what is set, so a random pattern and a working machine with the same number of specified bits count equally. Reversible manipulation and measurement, which have no thermal floor \citep{bennett1982thermodynamics}, are not counted. We do not model the fraction of power a civilization devotes to control operations, although that fraction separates civilizations at the same $K$ and $B$. By our convention an operation belongs to the civilization that specified its outcome, and the operations of an organism executing its own inherited genome do not, so selective breeding makes the boundary soft. Serial depth does not appear in $\Omega$. A Type III civilization on the thermal floor at 3\,K can perform $3.5\times10^{58}$ operations per second, but light takes $1.6\times10^{12}$\,s to cross its radiator, so the operations are nearly independent of one another. We also leave out the energy and mass of the machines and the heat removal that limits how densely operations can be packed. The count for a given artefact depends, through logarithmic factors, on the length at which its states are specified and the ensemble from which each is chosen. In the scale, Kardashev's $P$ is a budget for control operations, whereas Kardashev defined $P$ as the power of a transmitter. The power of another civilization can in principle be inferred from its waste heat, which Carrigan~\citep{carrigan2009iras} and Wright \textit{et al.}~\citep{wright2014ghat1} searched for in the infrared, and the temperature $T$ of that heat fixes the part $\tfrac{1}{10}\log_{10}(T_0/T)$ of its maturity gap. The heat alone determines neither the energy per operation of its devices nor its efficiency index $B$, since these also need the rate $\Omega$ of its operations, with $\eps = P/\Omega$. The rows of Table~\ref{tab:civ} for civilizations other than humanity use assumed values of $P$ and $\eps$.

\section{Conclusions}
\label{sec:conclusions}

This paper has introduced a technological capability scale built from the Kardashev and Barrow scales, with the index $\Xi = K + B$ and the quantity $\Omega = P/\eps$, the number of control operations a civilization can perform per second. Here $K$ is the Kardashev index of the power a civilization commands, which is the ceiling of the scale, and $B$ is an efficiency index of the energy it spends per control operation, measured against the thermal floor $\kB T_0\ln 2$ of the cosmic microwave background. For heat radiated into the present sky, $B \le 0$ for every operation that resets a thermally random degree of freedom, so $\Xi \le K$. At every level the least cost of such an operation is $\kB T\ln 2$ per bit at the temperature at which heat is rejected, so humanity's gains from working at smaller scales appear entirely as a narrowing of its maturity gap, measured against the same floor it had before 1950. The model values for adding a nucleon or creating a particle lie 11 and 15 orders above the floor at $T_0$. Miniaturization accounts for around 5 of the 11.7 orders by which switching energy has fallen. At the machines' own power, $\Xi$ rose by 1.48 from ENIAC to El~Capitan. In each of the four classes, namely processing, storage, transfer and construction, humanity's capability has grown mainly through $B$, which supplied 82\% to 92\% of the gain since the first documented cost, or for construction since the Jacquard loom of 1804. The cheapest processing cost 2 to 12\,J per operation from the abacus to ENIAC, so before 1946 humanity's capability for processing grew mainly with its power.

For humanity's logic, closing the maturity gap would raise $\Xi$ by 0.78, nearly three times the gain from reaching Type I. Of this, 0.57 comes from the devices and the rest, a supremum, from the temperature of the sink. In primary energy, the enzymes humanity directs lie about half as far from the thermal floor at 300\,K as its logic gates. Operations that reset thermally random degrees of freedom of matter cannot pass below the floor by reversible design, so once they reach the floor at their rejection temperature their budget grows only with power, with colder sinks or with new sinks for entropy. Copying a design into matter prepared in a known state has no such floor, so the distance of humanity's enzymes and fabrication plants from $\kB T\ln 2$ is not a maturity gap. The levels below the atom contribute mainly as sources of power and as dense storage. At a fixed rate of fuel consumption, local conservation of energy bounds the gain in power over combustion at nearly one Kardashev type. A programmable quantum field at this bound, operating arbitrarily close to the thermal floor at $T_0$, would approach $\Xi = K_{\mathrm{f}}$, the bound set by its rate of fuel consumption. The ultimate technology would dissipate the conjectured largest power $c^5/4G$ of one system into an uncharged, non-rotating black hole near its largest mass in de Sitter space, which places it at the corner of the $(K, B)$ plane at $\Xi = 7.58$.

Open questions include the least dissipation of a reliable operation at a given speed, the fraction of power a civilization devotes to control operations, an estimator of $B$ from observations that reveal the rate of operations of another civilization, and which physics, if any, lies between the electroweak and Planck scales.

\exhyphenpenalty=10000
\bibliographystyle{RS}
\bibliography{refs}

\end{document}